\documentclass[11pt,a4paper]{article}
\usepackage{amsmath,amssymb,amsfonts,a4wide,graphicx,bm,times}
\usepackage[colorlinks=true,linkcolor=blue, citecolor=blue,
urlcolor=blue]{hyperref} \input{epsf}
\numberwithin{equation}{section}
\makeatletter \let\old@startsection=\@startsection
\renewcommand{\@startsection}[6]
{\old@startsection{#1}{#2}{#3}{#4}{#5}{#6\mathversion{bold}}}
\makeatother

 \def\bu{{(2)}}
\def\ci{ {(1)}}
\def\nci{{n^{_{\ci }}}}
\def\nbu{{n^{_\bu}}}
\def\mci{\lambda_{^{\ci}}}
\def\mbu{\lambda_{^\bu}}
\def\naa{n^{_{(\a)}}}
\def\maa{\lambda_{^{(\a)}}}
\def\wci{\lambda_{^{\ci}}}
\def\wbu{\lambda_{^\bu}}
\def\waa{\lambda_{^{(\a)}}}
\def\aa{{(\a)}}
\def\defeq{\stackrel{\text{def}}=}
 \def\DJS{DJS\ } 
\newcommand\re[1]{({\ref{#1}})}
\def\be{\begin{eqnarray}}
    \def\ee{\end{eqnarray}}
    \def\no{\nonumber}
    \def\la{\label}
    \def\l {\lambda}
\def\({\left(} \def\){\right)} 
\def\<{\left\langle\,} 
\def\>{\, \right\rangle} 
\def\[{\left[}
 \def\]{\right]} 
\def\tr{{\rm   tr} }
    \def\hf{ {\textstyle{1\over 2}} }
\def\CO{{ \mathcal{ O} }}
  \def\p{\partial}
  \def\a{\alpha}
 \def\b{\beta}
  \def\g{\gamma}
  \def\e{\epsilon}
  
  \def\t{\tau}
  \def\th{\theta}
  \def\G{\Gamma}
      
\def\CA{{\cal A}} 
 \def\AA{{\cal A}}
 \def\BB{{\cal B}}
 \def\CC{{\cal C}}
 
 \def\reg{{\rm reg}}
 \def\O{\Omega} 

\begin{document}

\thispagestyle{empty}

\begin{flushright}
YITP-09-56 \\
IPhT-t09/186\\
\end{flushright}

\vspace{1cm}
\setcounter{footnote}{0}

\begin{center}

 {\Large\bf Boundary transitions of the   {\bf\it O(n)} model  on a dynamical  lattice }

\vspace{20mm} 

Jean-Emile Bourgine$^\ast$, Kazuo Hosomichi$^\star$ and Ivan
Kostov$^\ast$\footnote{\it Associate member of the Institute for
Nuclear Research and Nuclear Energy, Bulgarian Academy of Sciences, 72
Tsarigradsko Chauss\'ee, 1784 Sofia, Bulgaria}\\[7mm]

{\it $^\ast$ Institut de Physique Th\'eorique, CNRS-URA 2306 \\
	     C.E.A.-Saclay, \\
	     F-91191 Gif-sur-Yvette, France \\[5mm]

    $^\star$ Yukawa Institute for Theoretical Physics, Kyoto
    University \\
	     Kyoto 606-8502, Japan }\\
 
\end{center}

\vskip9mm

\vskip18mm

\noindent{ We study the anisotropic boundary conditions for the dilute
$O(n)$ loop model with the methods of 2D quantum gravity.  We solve
the problem exactly  on a dynamical lattice using the correspondence 
with a large $N$ matrix model.  We formulate the disk two-point functions 
with ordinary and anisotropic boundary conditions as loop correlators 
in the matrix model.   We derive the loop equations for these correlators 
and find their explicit solution in the scaling limit.   
Our solution reproduces the boundary phase diagram 
and the boundary critical exponents obtained recently by Dubail, 
Jacobsen and Saleur, except for the cusp at the isotropic special 
transition point.  Moreover, our solution describes the bulk and the boundary 
deformations away from the anisotropic special  transitions.  In particular it 
shows how the anisotropic special boundary conditions are deformed by 
the bulk  thermal flow towards the dense phase.
  }

\newpage

\setcounter{page}{1}
  
%&&&&&&&&&&&&&&&&&&&&&&&&&&&&&&&&&
\section{Introduction}
%&&&&&&&&&&&&&&&&&&&&&&&&&&&&&&&&&
\label{sec:Introduction}

The boundary critical phenomena appear in a large spectrum of
disciplines of the contemporary theoretical physics, from solid
state physics to string theory.  The most interesting situation is
when the boundary degrees of freedom enjoy a smaller symmetry than
those in the bulk.  In this case one speaks of surface anisotropy.
The D-branes in string theory are perhaps the most studied example of
such anisotropic surface behavior.  Another example is provided by the
ferromagnets with surface exchange anisotropy, which can lead to a
critical and multi-critical anisotropic surface transitions.  An
interesting but difficult task is to study the interplay of surface
and the bulk transitions and the related multi-scaling regimes.

 For uniaxial ferromagnets as the Ising model, there are four
 different classes of surface transitions: the ordinary,
 extraordinary, surface, and special transitions \cite{Diehl}.  This
 classification makes sense also for spin systems with continuous
 $O(n)$ symmetry.  It was predicted by Diehl and Eisenriegler
 \cite{DiEi}, using the $\varepsilon$-expansion and renormalization
 group methods, that the effects of surface anisotropy can be relevant
 near the special transitions of the $d$-dimensional $O(n)$ model.
 These effects lead to `anisotropic special transitions' with
 different critical exponents.
 
 Recently, an exact solution of the problem for the $2$-dimensional
 $O(n)$ model was presented by Dubail, Jacobsen and Saleur
 \cite{DJHS2} using its formulation as a loop model
 \cite{Domany:1981fg, Nienhuis:1982fx}.  Using an elaborate mixture of
 Coulomb gas, algebraic and Thermodynamic Bethe Ansatz techniques, the
 authors of \cite{DJHS2} confirmed for the dilute $O(n)$ model the
 phase diagram suggested in \cite{DiEi} and determined the exact
 scaling exponents of the boundary operators.  A review of the results
 obtained in \cite{DJHS2}, which is accessible for wider audience, can
 be found in \cite{Dubail:2009kx}.  Their works extended the techniques
 developped by Jacobsen and Saleur \cite{Jacobsen:2006bn, JS2} for the
 dense phase of the $O(n)$ loop model.

In this paper we examine the bulk and the boundary deformations away
from the anisotropic special transitions in the two-dimensional $O(n)$
model.  In particular, we address the question how the anisotropic
boundary transitions are influenced by the bulk deformation which
relates the dilute and the dense phases of the $O(n)$ model.  To make
the problem solvable, we put the model on a dynamical lattice.  This
procedure is sometimes called `coupling to 2D gravity' \cite{GDFZ,
GM}.  The sum over lattices erases the dependence of the correlation
functions on the coordinates, so they become `correlation numbers'.
Yet the statistical model coupled to gravity contains all the
essential information about the critical behavior of the original
model such as the qualitative phase diagram and the conformal weights of
the scaling operators.  When the model is coupled to gravity, the bulk
and boundary flows, originally driven by relevant operators, becomes
marginal, the Liouville dressing completing the conformal weights to
one.  This necessitates a different interpretation of the flows.  The
UV and the IR limits are explored by taking respectively large and
small values of the bulk and boundary cosmological constants.  Our
method of solution is based on the mapping to the $O(n)$ matrix model
\cite{Kostov:1988fy, GK} and on the techniques developed in
\cite{Kostov:2007jj, Bourgine:2008pg}.  Using the Ward identities of
the matrix model, we were able to evaluate the two-point functions of
the boundary changing operators for finite bulk and boundary
deformations away from the anisotropic special transitions.

The paper is structured as follows.  In Sect.  \ref{sec:O(n)flat} we
summarize the known results about the boundary transitions in the
$O(n)$ loop model.  In Sect.  \ref{sec:O(n)def} we write down the
partition function of the boundary $O(n)$ model on a dynamical
lattice.  In particular, we give a microscopic definition of the
anisotropic boundary conditions on an arbitrary planar graph.  In
Sect.  \ref{sec:Matrix} we reformulate the problem in terms of the
$O(n)$ matrix model.  We construct the matrix model loop observables
that correspond to the disk two-point functions with ordinary and
anisotropic special boundary conditions.  In Sect.  \ref{sec:Loop} we
write a set of Ward identities (loop equations) for these loop
observables, leaving the derivation to Appendix \ref{sec:Derivation}.
We are eventually interested in the scaling limit, where the volumes
of the bulk and the boundary of the planar graph diverge.  This limit
corresponds to tuning the bulk and the boundary cosmological constants
to their critical values.  In Sect.  \ref{sec:Scalimit} we write the
loop equations in the scaling limit in the form of functional
equations.  From these functional equations we extract all the
information about the bulk and the boundary flows.  In particular, we
obtain the phase diagram for the boundary transitions, which is
qualitatively the same as the one suggested in \cite{DJHS2,
Dubail:2009kx}, apart of the fact that we do not observe a cusp near
the special point.  In Sect.  \ref{sec:Spectr} we derive the conformal
weights of the boundary changing operators.  All our results
concerning the critical exponents coincide with those of \cite{DJHS2,
Dubail:2009kx}.  In Sect.  \ref{sec:Solution} we find the explicit
solution of the loop equations in the limit of infinitely large planar
graph.  The solution represents a scaling function of the coupling for
the bulk and boundary perturbations.  The endpoints of the bulk and
the boundary flows can be found by taking different limits of this
general solution.  The boundary flows relate the anisotropic special
transition with the ordinary or with the extraordinary transition,
depending on the sign of the perturbation.  The bulk flow relates an
anisotropic boundary condition in the dilute phase with another
anisotropic boundary condition in the dense phase.  For the rational
values of the central charge, the boundary conditions associated with
the endpoints of the bulk flow match with those predicted in the
recent paper \cite{Fredenhagen:2009sw} using perturbative RG
techniques.

   \section{The boundary $O(n)$ model on a regular lattice: a summary}

\label{sec:O(n)flat}

%%%%%%%%%%%%%%%%%%%%%%%%%%%%

The $O(n)$ model \cite{Domany:1981fg} is one of the most studied
statistical models.  For the definition of the partition function see
Sect.  \ref{sec:O(n)def}.  The model has an equivalent description in
terms of a gas of self- and mutually avoiding loops with fugacity $n$.
The partition function of the loop gas depends on the temperature
coupling $T$ which controls the length of the loops:
\be\la{loopgaz}
Z_{O(n)} = \sum_{\text{loops}} n^{\# [\text{loops}]}\({1/ T}\)^{\#[\text{links occupied by loops}]} .
\ee
In the loop gas formulation of the $O(n)$ model, the number of flavors
$n$ can be given any real value.  The model has a continuum transition
if the number of flavors is in the interval $[-2,2]$.  Depending on
the temperature coupling $T$
the model has two critical phases, the dense and the dilute phases.
At large $T$ the loops are small and the model has no long range
correlations.  The dilute phase is achieved at the critical
temperature \cite{Nienhuis:1984wm, NienhIJMP}
\be
\la{Tchoney}
T_c= \sqrt{2+\sqrt{2-n}}
\ee
for which the length of the loops diverges.  If we adopt for the
number of flavors the standard parametrization
\begin{equation}
 \la{paramn} n=2\cos(\pi\theta), \qquad 0<\theta <1,
\end{equation}
then at the critical bulk temperature the $O(n)$ model is described by
(in general) a non-rational CFT with central charge
\be\la{cdilute} c_{_{\text{dilute}}}= 1- 6 \, {\th^2\over 1+\th}.  \ee
The primary operators $\Phi_{r,s}$ in such a non-rational CFT can be
classified according to the generalized Kac table for the conformal
weights,
\be\la{KacTdil} h_{rs}= {(r g - s )^2 - (g-1)^2\over 4 g}, \qquad g =
1+\th.  \ee
Unlike the rational CFTs, here the numbers $r$ and $s$ can take
non-integer values.

When $T<T_c $, the loops condense and fill almost all space.  This
critical phase is known as the dense phase of the loop gas.  The dense
phase of the $O(n)$ is described by a CFT with lower value of the
central charge,
\be\la{cdense} c_{_{\text{dense}}}= 1- 6 \, {\th^2\over 1-\th}.  \ee
The generalized Kac table for the dense phase is
\be\la{KacTden} h_{rs}= {(r \tilde g - s )^2 - (\tilde g-1)^2\over 4
\tilde g}, \qquad \tilde g = {1\over 1-\th}.  \ee
Most of the exact results for the dense and the dilute phases of the
$O(n)$ model were obtained by mapping to Coulomb gas
\cite{Nienhuis:1982fx, Nienhuis:1984wm}.

The boundary $O(n)$ model was originally studied for the so called
{\it ordinary boundary condition}, where the loops avoid the boundary
as they avoid themselves.  The ordinary boundary condition is also
referred as {\it Neumann boundary condition} because the measure for
the boundary spins is free.  The boundary scaling dimensions of the
$L$-leg operators $S_L$, realized as sources of $L$ open lines, were
conjectured for the ordinary boundary condition in
\cite{Duplantier:1986zza} and then derived in \cite{SalBau}:
\be (Ord|S_L|Ord)\ \to \Phi_{1+L, 1}^B \qquad(\text{dilute
phase})\no\\
(Ord|S_L|Ord)\ \to \Phi_{1, 1+L} ^B \qquad(\text{dense phase}).
\ee
Another obvious boundary condition is the {\it fixed}, or {\it Dirichlet}, boundary
condition, which
allows, besides the closed loops, open lines that end at the boundary
\cite{Kazakov:1991pt}.\footnote{ In the papers \cite{Kazakov:1991pt,
Kbliou, KPS} the loop gas was considered in the context of the SOS
model, for which the Dirichlet and Neumann boundary conditions have
the opposite meaning.} The dimensions of the $L$-leg boundary
operators with Dirichlet and Neumann boundary conditions were computed
in \cite{Kbliou, KPS} by coupling the model to 2D gravity and then
using the KPZ scaling relation \cite{KPZ, DDK}:
\be (Ord|S_L|Dir)\ \to \Phi_{ 1/2+L , 0} ^B \qquad(\text{dilute
phase})\no\\
(Ord|S_L|Dir)\ \to \Phi_{0, 1/2+L}^B \qquad(\text{dense phase}).  \ee
From the perspective of the boundary CFT, these operators are obtained
as the result of the fusion of the $L$-leg operator and a
boundary-condition-changing (BCC) operator, introduced in
\cite{Cardy:1984bb}, which transforms the ordinary into Dirichlet
boundary condition.

Recently it was discovered that the $O(n)$ loop model can exhibit
unexpectedly rich boundary critical behavior.  Jacobsen and Saleur
\cite{Jacobsen:2006bn, JS2} constructed a continuum of conformal
boundary conditions for the dense phase of the $O(n)$ loop model.  The
{\it Jacobsen-Saleur boundary condition}, which we denote shortly by
JS, prescribes that the loops that touch the boundary at least once
are taken with fugacity $y\ne n$, while the loops that do not touch
the boundary are counted with fugacity $n$.  The boundary parameter
$y$ can take any real value.  The JS boundary conditions contain as
particular cases the ordinary ($y=n$) and fixed ($y=1$) boundary
conditions for the $O(n)$ spins.  Jacobsen and Saleur conjectured the
spectrum and the conformal dimensions of the $L$-leg boundary
operators separating the ordinary and the JS boundary conditions.
These conformal dimensions were subsequently verified on the model
coupled to 2D gravity in \cite{Kostov:2007jj, Bourgine:2008pg}.
If $y$ is parametrized in the `physical' interval $0\le y\le n$ as
\be
\la{paramndense}
y = {\sin \pi (r+1)\th\over \sin \pi r\th} 
\qquad (1\le r\le 1/\th-1),
\ee
then the BCC operator transforming
 the ordinary into JS boundary
condition is identified as the diagonal operator $\Phi_{r,r}$.  Note
that $r$ here is not necessarily integer or even rational.  The
$L$-leg operators with $L\ge 1$ fall into two types.  The operator $S^-_L$
creates open lines which avoid the JS boundary.  The operator $S^+_L$
creates open lines that can touch the JS boundary without restriction.
The two types of $L$-leg boundary operators are identified as
\be (Ord|S_L^\pm |JS)\ \to \Phi^B _{r, r\pm L} \qquad(\text{dense
phase}).  \ee
The general case of two different JS boundary conditions with boundary
parameters $y_1$ and $y_2$ was considered for regular and dynamical
lattices respectively in \cite{DJS} and \cite{Bourgine:2009hv}.

The JS boundary condition was subsequently adapted to the dilute phase
by Dubail, Jacobsen and Saleur \cite{DJHS2}.  The authors of
\cite{DJHS2} considered the loop gas analog of the anisotropic
boundary interaction studied previously by Diehl and Eisenriegler
\cite{DiEi}, which breaks the symmetry as
 \be O(n)\to O(\nci)\times O(\nbu), \quad \nci+\nbu=n.  \ee
The boundary interaction depends on two coupling constants, $\wci$ and
$\wbu$, associated with the two unbroken subgroups.  In terms of the
loop gas, the anisotropic boundary interaction is defined by
introducing loops of two colors, $\ci$ and $\bu$, having fugacities
respectively $\nci$ and $\nbu$.  Each time when a loop of color $(\a)$
touches the JS boundary, it acquires an extra factor $\waa$.  We will
call this boundary condition {\it dilute Jacobsen-Saleur boundary
condition}, or shortly \DJS, after the authors of \cite{DJHS2}.

Let us summarize the qualitative picture of the surface critical
behavior proposed in \cite{DJHS2}.  Consider first the isotropic
direction $\wci=\wbu =\l $.  In the dilute phase one distinguishes three
different kinds of critical surface behavior: ordinary, extraordinary
and special.  When $\l=0$, the loops in the bulk almost never touch the
boundary.  This is the ordinary boundary condition.  When
$w\to\infty$, the most probable loop configurations are those with one
loop adsorbed along the boundary, which prevents the other loops in
the bulk to touch the boundary.  The adsorbed loop plays the role of a
boundary with ordinary boundary condition.  This is the extraordinary
transition.  In terms of the $O(n)$ spins, ordinary and the
extraordinary boundary conditions describe respectively disordered and
ordered boundary spins.\footnote{The spontaneous ordering on the
boundary does not contradict the Mermin-Wagner theorem.  In the
interval $1<n<2$ the target space, the $(n-1)$-dimensional sphere, has
negative curvature and thus resembles a non-compact space.} The
ordinary and the extraordinary boundary conditions describe the same
continuous theory except for a reshuffling of the boundary operators.
The $L$-leg operator with ordinary/extraordinary boundary conditions
will look, when $L\ge 1$, as the $(L-1)$-leg operator with
ordinary/ordinary boundary conditions, because its rightmost leg will
be adsorbed by the boundary.  The $0$-leg operator with
ordinary/extraordinary boundary condition will look like the $1$-leg
operator with ordinary/ordinary boundary condition because one of the
vacuum loops will be partially adsorbed by the extraordinary boundary
and the part which is not adsorbed will look as an open line
connecting the endpoints of the extraordinary boundary.

The ordinary and the extraordinary boundary conditions are separated
by a special transition, which happens at some $\l=\l_c $ and describes
a conformal boundary condition.  For the honeycomb lattice the special
point is known \cite{Batyu} to be at
\be
\l_c= (2-n)^{-1/2} \ T_c^2 \qquad (\text{honeycomb lattice}).
\la{mspflat}
\ee
At the special point the loops touch the boundary without being
completely adsorbed by it.  The special transition exists only in the
dilute phase, because in the dense phase the loops already almost
surely touch the boundary.  The only effect of having small or large
$\l $  is the reshuffling of the spectrum of $L$-leg operators, which
happens in the same way as in the dilute phase.

In the anisotropic case, $\wci \ne \wbu$, there are again three
possible transitions: ordinary, extraordinary and special.  When $\wci
$ and $\wbu$ are small, we have the same ordinary boundary condition
as in the isotropic case.  In the opposite limit, where $\wci $ and
$\wbu$ are both large, the boundary spins become ordered in two
different ways, depending on which of the two couplings prevails.  If
$\wci >\wbu $, the $\ci$-type components order, while the $\bu$-type
components remain desordered, and {\it vice versa}.  The $(\mci ,
\mbu)$ plane is thus divided into three domains, characterized by
disordered, $\ci$-ordered and $\bu$-disordered, and $\bu$-ordered and
$\ci$-disordered, which we denote respectively by ${ Ord}$, $Ext_\ci$
and $Ext_\bu$.  The domains $Ext_\ci$ and $Ext_\bu$ are separated by
the isotropic line starting at the special point and going to
infinity.  This is a line of first order transitions because crossing
it switches from one ground state to the other.  The remaining two
critical lines, $Ord/Ext_\ci$ and $Ord/Ext_\bu$, are the lines of the
two {\it anisotropic special transitions}, $AS_\ci$ and $AS_\bu$.  It
was argued in \cite{DiEi, DJHS2, Dubail:2009kx}, using scaling
arguments, that the lines $AS_\ci$ and $AS_\bu$ join at the point $Sp=
\(\l_c,\l_c\)$ in a cusp-like shape.  The model was solved in
\cite{DJHS2} for a particular point on $AS_\ci$:
\be
\wci &=& 1 + {1- \nbu + \sqrt{1-\nci\nbu}\over \sqrt{2-n}}, \no\\
\wbu& =& 1 + {1- \nci - \sqrt{1-\nci\nbu}\over \sqrt{2-n}} \, .
\ee

In terms of the loop gas expansion, the anisotropic special
transitions are obtained by critically enhancing the interaction with
the boundary of the loops of color $\ci$ or $\bu$.  The boundary CFT's
describing the transitions $AS_\ci$ and $AS_\bu$ were identified in
\cite{DJHS2}.  A convenient parametrization of $\nci$ and $\nbu$ on
the real axis is\footnote{The correspondence between our  notations 
and the notations used in \cite{Dubail:2009kx} 
is $\{\nci,  \nbu, \, r , \, \wci, \wbu\}_{\text{here}}=
\{ n_2,n_1, \, r,\,  w_2,w_1\}_{\text{there}}$.  }
\begin{equation}
\la{paramen} \nci={\sin[\pi(r -1)\theta]\over\sin[\pi r \theta]},
\quad \nbu={\sin[\pi(r +1)\theta]\over\sin[\pi r \theta]} \qquad \( 0<
r <{1/ \th}\)\, .
\end{equation}
The loop model has a statistical meaning only if both fugacities are
positive, which is the case when $1<r<1/\th-1$.  With the above
parametrization, the BCC operators $(AS_\ci
|Ord)$ and $(AS_\bu|Ord)$ are argued to be respectively $\Phi^B_{r,r}$
and $\Phi^B_{r, r+1}$.  More generally, one can consider the $L$-leg
boundary operators, $S_L^\ci$ and $S_L^\bu$, which create $L$ open
lines of color respectively $\ci$ and $\bu$.  The Kac labels of these
operators were determined in \cite{DJHS2} as follows,
 \be
 (Ord |S_L^{_{\ci}} |  AS_\ci)&\to\   \Phi_{r+L,r}^B,
 \quad
 (Ord |S_L^{_{\bu}} | AS_\ci)&\to \Phi_{r-L,r}^B,\no\\
 (Ord|S_L^{_{\ci}} | AS_\bu)&\ \to \Phi_{r+L,r+1}^B,
 \quad
 (Ord |S_L^{_{\bu}} | AS_\bu)&\to \Phi_{r-L,r+1}^B\, .
  \ee
In the vicinity of the special transitions the theory is argued to be
described by a perturbation of the boundary CFT by the boundary
operator $\Phi_{1,3}^B$ in the isotropic direction and by
$\Phi_{3,3}^B$ in the anisotropic direction.

\section{The boundary  $O(n)$ model  on a dynamical lattice }
%%%%%%%%%%%%%%%%%%%%%%%%%%%%%%%%%

\label{sec:O(n)def}

\subsection{Anisotropic boundary conditions for the $O(n)$ model on a
planar graph}

\begin{figure} 
\begin{center}
\includegraphics[width=145pt]{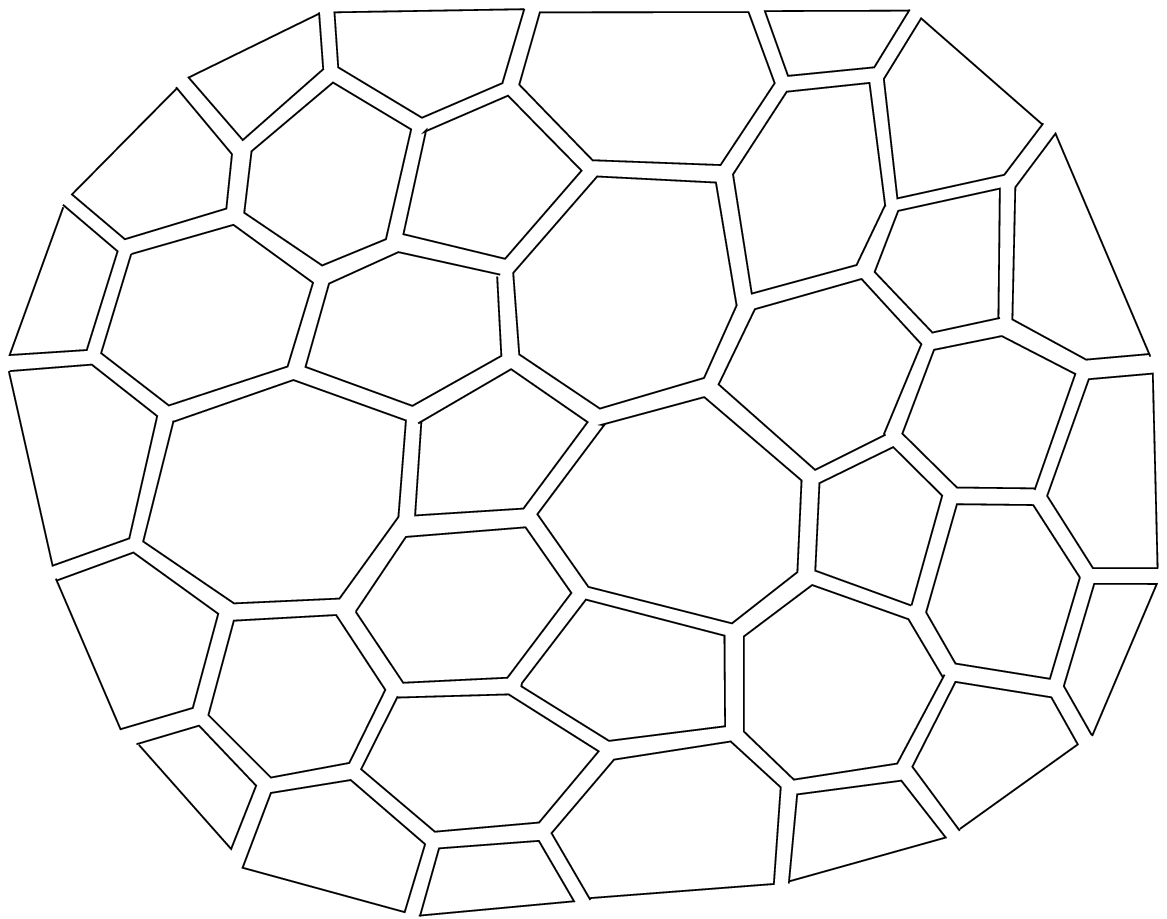}\hskip2cm
\includegraphics[width=145pt]{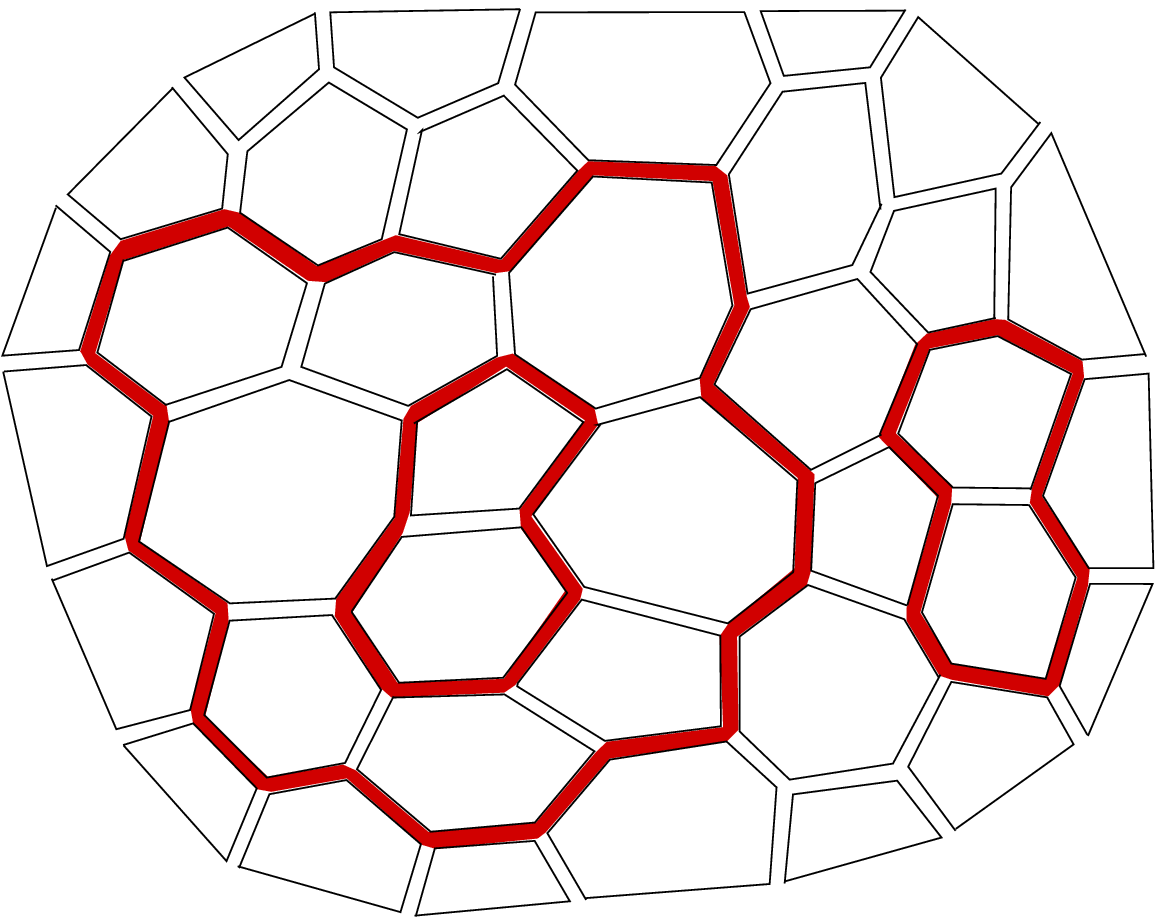} \\
a \hskip 7cm b
\end{center}
 \caption{\small a)~ A trivalent planar graph $\G$ with a boundary ~~
 b)~ A loop configuration on $\Gamma$ for the ordinary boundary
 condition.  The loops avoid the boundary as they avoid themselves}
\label{Diskabc}
\end{figure}

The two-dimensional $O(n)$ loop model, originally defined on the
honeycomb lattice \cite{Domany:1981fg}, can be also considered on a
honeycomb lattice with defects, such as the one shown in Fig.
\ref{Diskabc}a.  The lattice represents a trivalent planar graph
$\Gamma$.  We define the boundary $\p\Gamma$ of the graph by adding a
set of extra lines (the single lines in the figure) which turn the
original planar graph into a two-dimensional cellular complex.  The
local fluctuating variable is an $O(n)$ classical spin, that is an
$n$-component vector $\vec S(r)$ with unit norm, associated with each
vertex $r\in\Gamma$, including the vertices on the boundary
$\p\Gamma$.  The partition function of the $O(n)$ model on the graph
$\Gamma$ depends on the coupling $T$, called temperature, and is
defined as an integral over all classical spins,
\begin{equation}\label{partfOn}
Z_{_{O(n)}}(T;\Gamma)=\int\prod_{r\in\Gamma}[d\vec S(r)]
\prod_{\langle rr' \rangle}\Big(1+\frac1T\, \vec S(r)\cdot \vec S(r')
\Big)\,,
\end{equation}
where the product runs over the lines $\langle rr'\rangle$ of the
graph, excluding the lines along the boundary.  The $O(n)$-invariant
measure $[d\vec S]$ is normalized so that
\be
\int [d\vec S] \ S_a S_b = \delta_{a,b}.
\ee
The partition function (\ref{partfOn}) corresponds to the {\it
ordinary boundary condition}, in which there is no interaction along
the boundary.

Expanding the integrand as a sum of monomials, the partition function
can be written as a sum over all configurations of self-avoiding,
mutually-avoiding loops as the one shown in Fig.~\ref{Diskabc}b, each
counted with a factor of $n$:
\begin{equation}\label{looprON}
 Z_{_{O(n)}}(T;\Gamma)=\sum_{\text{loops on }\Gamma}\;
 T^{-\text{length }}\, n^{\#\text{loops}}.
\end{equation}
The temperature coupling $T$ controls the length of the loops.  The
advantage of the loop gas representation (\ref{looprON}) is that it
makes sense also for non-integer $n$.  In terms of loop gas, the
ordinary boundary condition, which we will denote by $Ord$, means that
the loops in the bulk avoid the boundary as they avoid the other loops
and themselves.

The {\it Dirichlet boundary
condition}, was originally defined for the dense phase of the
loop gas \cite{Kazakov:1991pt,Kbliou,KPS} and requires that an open
line starts at each point on the boundary.  The dilute version of this
boundary condition depends on an adjustable parameter, which controls
the number of the open lines.  In terms of the $O(n)$ spins the
Dirichlet boundary condition is obtained by switching on a constant
magnetic field $\vec B$ acting on the boundary spins.
 This modifies the integration measure in \re{partfOn} by a factor
 \be \label{Dirichl} \prod_{ r\in \p\Gamma} \Big(1+ \, \vec B\cdot
 \vec S (r) \Big)\, ,  \ee
  where the product goes over all boundary sites $r$.
 The loop expansion with this boundary measure contains open lines
 having both ends at the boundary, each weighted with a factor $\vec B^2$.

\begin{figure} 
\begin{center}
\includegraphics[width=290pt]{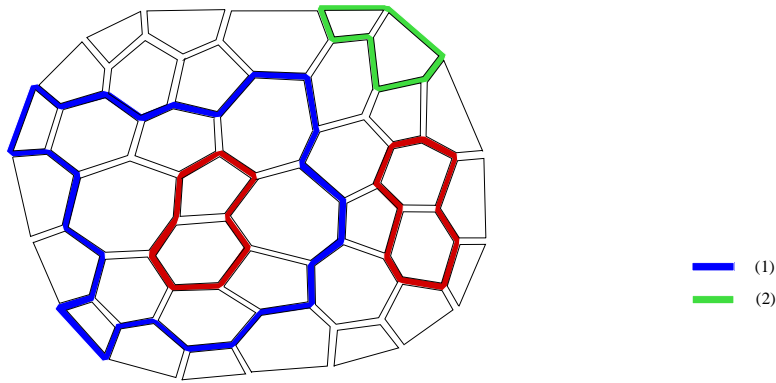} \\
\end{center}
 \caption{\small A loop configuration for the JS boundary condition.
 The loops in the bulk (in red) have fugacity $n$, while the loops
 that touch the boundary (in blue and green) have fugacities $\nci$ or
 $\nbu$ depending on their color.    }
\label{DiskJS}
\end{figure}

The {\it dilute anisotropic (DJS) boundary condition}    is defined  as follows. 
The $n$ components of the $O(n)$ spin  are split into two sets,
$\ci$ and $\bu$, containing respectively $\nci$ and $\nbu$ components,
with $\nci+\nbu=n$.  This leads to a decomposition of the $O(n)$ spin as
 \be\la{orthg}
 \vec S = \vec S_{^\ci}+\vec S_{^\bu},
 \quad \vec S_{^\ci}\cdot \vec S_{^\bu}=0.
 \ee
The   \DJS boundary condition 
is introduced by an extra factor  associated with the boundary links,
\begin{equation}\label{JSboundw2}
 \prod_{ \<r r'\!\>\in \p\Gamma} \Big(1+\sum_{\a=1,2} \waa\, \vec S_{^\aa}(r)\cdot
 \vec S_{^{\aa}}(r')   \Big)\, .
\end{equation}
This  boundary interaction  is invariant under the subgroup of independent rotations 
of $\vec S_\ci$ and  $\vec S_\bu$.   
The boundary term changes the loop expansion.  The loops are now
allowed to pass along the boundary links as shown in
Fig.~\ref{DiskJS}.  We have to introduce loops of two colors, $\ci$
and $\bu$, having fugacities respectively $\nci$ and $\nbu$.  A loop
of color $(\a)$ that visits ${\cal N}$ boundary links acquires an
additional weight factor $\maa^{\cal N}$.  For the loops that
do not touch the boundary, the contributions of the two colors sum up
to $\nci+\nbu=n$ and we obtain the same weight as with the ordinary
boundary condition.

\subsection{Coupling to 2D discrete gravity}

The disk partition function of the $O(n)$ model on a dynamical lattice
is defined as the expectation value of (\ref{partfOn}) in the ensemble
of all trivalent planar graphs $\Gamma$ with the topology of the disk.
The measure depends on two more couplings, $\bar\mu$ and $\bar\mu_B$,
called respectively bulk and  boundary cosmological
constants,\footnote{We used bars to distinguish from the bulk and
boundary cosmological constants in the continuum limit.} associated
with the volume $|\Gamma|= \#(\text{cells})$ and the boundary length
$|\partial\Gamma|=\#(\text{external lines})$.  The partition function
of the disk is a function of $\bar\mu$ and $\bar\mu_B$ and is defined
by
\begin{equation}\label{defPPG}
U(T,\bar\mu,\bar\mu_B) = \sum_{\Gamma\in\{\text{Disk}\}}
\frac1{|\partial\Gamma|} \ \({1\over \bar{\mu}}\)^{{|\Gamma|} }
\({1\over \bar\mu_B}\)^{{ |\partial\Gamma|}} Z_{_{O(n)}}(T;\Gamma).
\end{equation}

 \subsection{Two-point functions of the $L$-leg boundary operators}

Our aim is to evaluate the boundary two-point function of two $L$-leg
operators separating ordinary and anisotropic boundary conditions.
The $L$-leg operator $S_L$ is obtained by fusing $L$ spins with flavor
indices $a_1,\dots, a_n \in\{1,\dots , n\}$.  In terms of the loop
gas, the operator $S_L$ creates $L$ self and mutually avoiding open
lines.  We would like to exclude configurations where some of the
lines contract among themselves.  This can be achieved by taking the
antisymmetrized product
\begin{equation}\label{defSL}
 S_L  \sim \det_{L\times L}S_{a_i}(r_j) ,
\end{equation}
where $r_1,\dots, r_L$ are $L$ consecutive boundary vertices of the
planar graph $\G$, and we put the label $L$ instead of writing its
dependence on $a_1,\dots,a_L$ explicitly.  The two-point function of
the operator $S_L$ is evaluated as the partition function of the loop
gas in presence of $L$ open lines connecting the points $\{r_i\}$ and
$\{r'_i\}$.  The open lines are self- and mutually avoiding, and are
not allowed to intersect the vacuum loops.

Since the \DJS  boundary condition breaks the $O(n)$ symmetry  into
$O(\nci)\times O(\nbu)$, there are two inequivalent correlation
functions of the $L$-leg operators with $Ord/\DJS$  boundary
conditions.  Indeed, the insertion of $S_a$ has different effects
depending on whether $a$ belongs to the $O(\nci)$ or the $O(\nbu)$
sectors.  In the first case the open line created by $S_a$ acquires a
factor $\mci  $ each time it visites a boundary link.  In the second
case, the factor is $\mbu$.  Therefore the boundary spin operators
(\ref{defSL}) with $Ord/\DJS$  boundary conditions split into
two classes,
\begin{eqnarray}\label{SnJS}
  S_L^\ci &\sim & \det[S_{a_i}(r_j)]\,,\quad a_1,\dots,a_L\in \ci ,
  \nonumber\\
  S_L^\bu &\sim & \det[S_{a_i}(r_j)]\,,\quad a_1,\dots,a_L\in \bu .
\end{eqnarray}
We denote the corresponding boundary two-point functions respectively
by $D_L^\ci $ and $D_L^\bu$.

%%%%%%%%%%%%%%%%%%%%%%%%%%%%%%%%%%%%%%%%%%%%
\section{Mapping to the $O(n)$ matrix model}
%%%%%%%%%%%%%%%%%%%%%%%%%%%%%%%%%%%%%%%%%%%%
\label{sec:Matrix}

\begin{figure}
\begin{center}
\includegraphics[width=300pt]{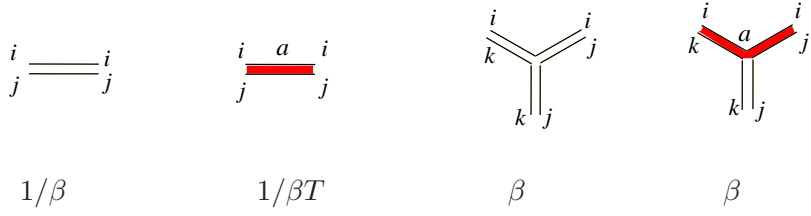}
\end{center}
\centerline{$1/\b$ \hskip 2.4cm $1/\b T$ \hskip 2.3cm $\b$ \hskip
2.5cm $\b$\qquad }
\caption{\small Feynman rules for the $O(n)$ matrix model }
\label{fig:feinm}
\vskip 0.5 cm
\end{figure}

The $O(n)$ matrix model \cite{Kostov:1988fy, GK} generates planar
graphs covered by loops in the same way as the one-matrix models
considered in the classical paper \cite{BIPZ} generate empty planar
graphs.  The model involves the hermitian $N\times N$ matrices ${\bf
M}$ and ${\bf Y}_a$, where the flavor index $a$ takes $n$ values.  The
partition function is given by an $O(n)$-invariant matrix integral
\begin{equation}\label{partfXY}
 {\cal Z}_N(T) \sim \int d{\bf M}\, d^n{\bf Y}\, e^{- \beta\,
 \text{tr}\left( \frac12{\bf M}^2+\frac T2\vec{\bf Y}^2 -\frac13{\bf
 M}^3-{\bf M}\vec{\bf Y}^2\right) }.
\end{equation}
This integral can be considered as the partition function of a
zero-dimensional QFT with Feynman rules given in Fig.~\ref{fig:feinm},
where we used the 't Hooft double-line notations.  The graphs made of
such double-lined propagators are known as fat graphs.  The `vacuum
energy' of the matrix model represents a sum over connected fat
graphs, which can be also considered as discretized two-dimensional
surfaces of all possible genera.  As the action is quadratic in the
matrices ${\bf Y}_a$, their propagators arrange in closed loops
carrying a flavor $a$.  The sum of all Feynman graphs with given
connectivity can be viewed as the sum over all configurations of self
and mutually avoiding loops on a given discretized surface.  The
weight of each loop is given by the product of factors $1/T$, one for
each link, and the number of flavors $n$.  We are interested in the
large $N$ limit
\begin{equation}\label{limitNc}
 N \to\infty,\qquad \beta/N= \bar\mu^2 ~~(\text{fixed})\, ,
\end{equation}
in which only fat   graphs of genus zero survive \cite{tHooft}.

The basic observable in the matrix model is the resolvent
\be
W(\bar \mu_B) 
=    \frac1\beta \<\text{tr} {1\over \bar\mu_B -   {\bf M} }\>\, ,
\ee
 evaluated in the ensemble (\ref{partfXY}).  The resolvent is the
 one-point function with ordinary boundary conditions and is related
 to the disk partition function by $ W= - \p_{\bar \mu_B} U$.

\begin{figure}
\begin{center}
\includegraphics[width=175pt]{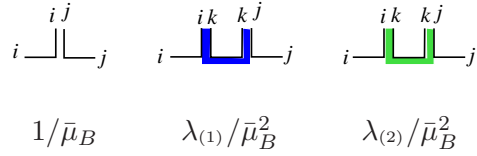}
\end{center}
\centerline{\hskip 2cm $1/\bar\mu_B$ \hskip 1cm $\wci/\bar\mu_B^2$
\hskip 1cm $\wbu/\bar\mu_B^2$\hskip 2cm }
\caption{\small The constituents of the \DJS boundary generated by
${\bf M}$, $\vec {\bf Y}_\ci^2$ and $\vec {\bf Y}_\bu^2$: a
non-occupied site, a sequence of two sites visited by a loop of color
$\ci$, and a sequence of two sites visited by a loop of color $\ci$}
\label{fig:feinmB}
\vskip 0.5 cm
\end{figure}

The one-point function  with Dirichlet boundary condition is 
obtained by adding a term $ \vec B\cdot \vec {\bf Y}$ which
expresses the coupling with the magnetic field on the boundary.
 This leads to a more-complicated resolvent
\be
R(\bar \mu_B, \vec B^2) = {1\over\b}
 \< \tr {1\over \bar  \mu_B -{\bf M} - \vec B\cdot \vec {\bf Y}}\>.
\ee

In order to include the anisotropic boundary conditions in this
scheme, we decompose the vector $\vec {\bf Y}$ into a sum of an
$\nci$-component vector $\vec {\bf Y}_{^\ci}$ and an $\nbu$-component
vector $\vec {\bf Y}_{^\bu}$ as in \re{orthg}:
\be \vec {\bf Y}= \vec
{\bf Y}_{^\ci}+\vec {\bf Y}_{^\bu}, \quad \vec {\bf Y}_{^\ci}\cdot \vec {\bf
Y}_{^\bu}=0.  \ee 
The one-point function with ordinary and \DJS boundary conditions is
given by the resolvent
\be\la{defH} H(\bar \mu_B,\wci,\wbu) = \frac1\beta \<\text{tr}
{1\over \bar \mu_B -  {\bf M} - {1\over \bar \mu_B  }
\sum_{\a=1,2 } \waa \vec {\bf Y}_\aa^2 }\, \>\, . \ee 
 The two extra terms are  the operators creating boundary links containing
segments of lines of type $\ci$ and $\bu$, as shown in
Fig.\ref{fig:feinmB}. Each such operator created two boundary 
sites,  hence the factor $1/\bar\mu_B$.

The matrix integral measure becomes singular at ${\bf M} = T/2$.  We
perform a linear change of the variables
\begin{equation}\la{shiftx}
 {\bf M}= T\big(\tfrac12+{\bf X}\big)\, ,
\end{equation}
which sends this singular point to ${\bf X}=0$.  After a suitable
rescaling of $\vec{\bf Y}$ and $\beta$, the matrix model partition
function takes the canonical form
\begin{equation}\label{partf}
{\cal Z}_N \sim \int d{\bf X}\, d^n{\bf Y}
e^{\beta\tr[-V({\bf X})+{\bf X}\vec{\bf Y}^2]}\,,
\end{equation}
where $V(x)$ is a cubic potential
\begin{equation}
V(x)= \sum_{j=0}^3 \frac{g_j}jx^j 
    = -\frac T3\Big(x+\tfrac12\Big)^3+\frac12\Big(x+\tfrac12\Big)^2.
\end{equation}
We introduce the spectral parameter $x$ which is related to the lattice
boundary cosmological constant $\bar\mu_B$ by
\begin{equation}
 \bar\mu_B = T\big(x+\tfrac12\big)\, .
\end{equation}
Now   the  one-point function with ordinary boundary condition is
\be\label{resW}
 W(x)= \frac1\beta\left\langle\text{tr}{\bf W}(x)\right\rangle , 
 \ee
 where the matrix
 \be
 {\bf W}(x)~\defeq~ \frac1{x-{\bf X}}\,.
\ee
creates a boundary segment with open ends.  In the following we will
call $x$ boundary cosmological constant.  We also redefine the
boundary couplings  $\waa$ in \re{defH} as
 \be\la{defx} \waa \to  {\maa \, \bar \mu_B} ,\quad \a=1,2\,.  \ee
Then the operator that creates a boundary segment with \DJS boundary
condition is
\begin{equation}\label{resWJSM}
 {\bf H}(y)~\defeq~ \frac1{y-{\bf X}-\mci \vec{\bf Y}_{^\ci} ^2 -\mbu
 \vec {\bf Y}_{^\bu}^2}\, .
\end{equation}

The boundary $L$-leg operators are represented by the antisymmetrized
products
\begin{equation}\label{SLmat}
 {\bf S}_L \defeq {\bf Y}_{a_1}{\bf Y}_{a_2}\cdots{\bf Y}_{a_L}\pm
 \text{permutations}\,.
\end{equation}
 The boundary two-point function of the $L$-leg operator with
 $Ord/Ord$ boundary conditions is given by the expectation value
\begin{equation}\label{defDL}
 D_L(x_1,x_2)~\defeq~\frac1\beta\left\langle \text{tr}\big[{\bf
 W}(x_1){\bf S}_L{\bf W}(x_2){\bf S}_L\big]\right\rangle.
\end{equation}
The role of the operators ${\bf W}(x_1)$ and ${\bf W}(x_2)$ is to
create the two boundary segments with boundary cosmological constants
respectively $x_1$ and $x_2$.  The two insertions ${\bf S}_L$ generate
$L$ open lines at the points separating the two segments.  It is
useful to extend this definition to the case $L=0$, assuming that
$S_0$ is the boundary identity operator.  In this simplest case the
expectation value (\ref{defDL}) is evaluated instantly as
\begin{equation}\label{DNNo}
 D_0(x_1,x_2)~=~\frac1\beta\left\langle \text{tr}\big[{\bf W}(x_1){\bf
 W}(x_2)\big]\right\rangle ~=~ \frac{W(x_2)-W(x_1)}{x_1-x_2}.
\end{equation}
The two-point functions \re{defDL} for $L\ge 1$ 
were computed   in \cite{Kbliou,
KPS}.

In the case of a \DJS boundary condition, the matrix model realization
of the two types of boundary $L$-leg operators is given by the
antisymmetrized products (\ref{SLmat}), with the restrictions on the
components as in (\ref{SnJS}).  The boundary two-point functions with
$Ord/\DJS$ boundary conditions are evaluated by the expectation values
\be 
\label{Btp0} 
D_0 (x,y) &=& \frac1\beta\Big\langle\text{tr}\big[
{\bf W}(x)  {\bf H}(y) \big]\Big\rangle
\ee
and for $L\ge 1$,
\begin{eqnarray}
\label{BtpfM} D_L^\ci (x,y) &=& \frac1\beta\Big\langle\text{tr}\big[
{\bf W}(x){\bf S}_L^\ci {\bf H}(y){\bf S}_L^\ci \big]\Big\rangle, \\
\label{BtpfMP} D_L^\bu(x,y) &=& \frac1\beta\Big\langle\text{tr}\big[
 {\bf W}(x){\bf S}_L^\bu{\bf H}(y){\bf S}_L^\bu\big]\Big\rangle .
\end{eqnarray}
Apart of the \DJS boundary parameter $\nci $ and the boundary couplings
$\mci  ,\mbu$ for the second segment, they depend on the boundary
cosmological constants $x$ and $y$ associated with the
two segments of the boundary.

%%%%%%%%%%%%%%%%%%%%%%%
\section{Loop equations}
%%%%%%%%%%%%%%%%%%%%%%%
\label{sec:Loop}

Our goal is to evaluate the two-point functions \re{BtpfM} in the
continuum limit, when the area and the boundary length of the disk are
very large.  They will be obtained as solution of a set of Ward
identities, called loop equations, which follow from the translational
invariance of the integration measure in \re{partfXY}, and in which
the $\nci$ enters as a parameter.  The solutions of the loop equations
are analytic functions of $\nci$ which can take any
real value.  We will restrict our analysis to the `physical' case
$0\le\nci\le n$, when the correlation functions have  good
statistical limit.  Here we summarize the loop equations which will be
extensively studied in following sections.  The proofs are given in
  Appendix \ref{sec:Derivation}.
 
%%%%%%%%%%%%%%%%%%%%%%%%%%%%%%%%%%%
%%%%%%%%%%%
\subsection{Loop equation for the resolvent  }
%%%%%%%%%%%%%%%%%%%%%%%%%%%%%%%%%%%
%%%%%%%%%%%

The loop equation for the resolvent is known \cite{Kostov:1991cg}, but
we nevertheless recall it here in order to set up a self-contained
description of the method.  The resolvent $W(x)$ splits into a
singular part $w(x)$ and a polynomial $W_\reg(x)$:
\be \la{WregWsin} 
 W(x) \defeq W_\reg(x) +   w(x) \, .
 \ee
 The regular part  is given by
 \be W_\reg(x)&=& \frac{2 V'(x)-n V'(-x)}{4-n^2} =- a_0-a_1x-a_2x^2
 \,,
 \label{defGx}
 \no\\ a_0&=& - { g_1\over 2+n} =\frac{T-2}{4(2+n)}, \no\\
\quad a_1&=& -{g_2\over 2-n}=\frac{T-1}{2-n}, \no\\
\quad a_2&=& -{g_3\over 2+n}=\frac T{2+n}.
\end{eqnarray}
The function $w(x)$ satisfies a quadratic identity
\begin{equation}\label{funceqW}
 w^2(x)+w^2(-x)+nw(x)w(-x)= A+Bx^2+Cx^4\,.
\end{equation}
The coefficients $A,B,C$ as functions of $ T$, $\bar\mu$ and $W_1=
\langle\text{tr}X\rangle$ can be evaluated by substituting the
large-$x$ asymptotics
\begin{equation}\label{asymG}
 w(x) = - W_\reg(x) +\frac{\bar\mu^{-2}}x
 +\frac{\langle\text{tr}X\rangle}{\beta x^2}+O(x^{-3})
\end{equation}
in \re{funceqW}.  The solution of the loop equation \re{funceqW} with
the asymptotics \re{asymG} is given by a meromorphic function with a
single cut $[a,b]$ on the first sheet, with $a<b<0$.  This equation
can be solved by an elliptic parametrization and the solution is
expressed in terms of Jacobi theta functions \cite{EK}.

%%%%%%%%%%%%%%%%%%%%%%%%%%%%%%%%%%%%% 
\subsection{Loop equations for the boundary two-point functions
with Ord/\DJS boundary conditions}
%%%%%%%%%%%%%%%%%%%%%%%%%%%%%%%%%%%%%% 

The two-point correlators \re{defDL} with ordinary boundary conditions
are known to satisfy the integral recurrence equations \cite{Kbliou,
KPS}
\be\la{recDL} D_{L+1} = W\star D_L\, .\ee
The ``$\star$-product'' is defined for any pair of meromorphic
functions, analytic in the right half plane $\text{Re}(x)\ge0$ and
vanishing at infinity,
\begin{equation}\label{defstar}
 [f\star g](x)~\defeq~ -\oint\limits_{{\cal C}_-}\frac{dx'}{2\pi i}
 \frac{f(x)-f(x')}{x-x'}g(-x'),
\end{equation}
with the contour ${\cal C}_-$ encircling the left half plane
$\text{Re}x<0$.  These equations actually hold for a more general set
of two-point correlators, which have ordinary boundary condition on one
segment and an arbitrary boundary condition on the other segment
\cite{Kostov:2007jj}.  Thus the boundary two-point functions
(\ref{BtpfM}) and (\ref{BtpfMP}) for $L\ge 1$ satisfy the same
recurrence equations
\begin{eqnarray}
 D_{L+1}^\ci &=& W\star D_L^\ci \,, \nonumber \\
 D_{L+1}^\bu &=& W\star D_L^\bu \, .
\label{recestar}
\end{eqnarray}
Using the recurrence relation, the correlation functions of the
$L$-leg operators can be obtained recursively from those of the
one-leg operators $D_1^\ci$ and $D_1^\bu$.

 The  correlator  $D_0$, defined by \re{Btp0}, and the 
 correlators $D_1^\ci$ and $D_1^\bu$, which we normalize as
 \be D_1^\aa  (x,y) &=& \frac{1}{\beta \naa }  \sum_{a}
 \left\langle\text{tr}\big[ {\bf W}(x){\bf Y}^\aa_a {\bf H}(y){\bf Y}^\aa_a
 \big]\right\rangle \qquad(\a=1,2),
 \ee
can be determined by the following pair of bilinear functional
equations, derived in Appendix \ref{sec:Derivation}.  In order to
shorten the expressions, here and below we use the shorthand notation
\be \overline{F(x)} \defeq F(-x)\, .
\ee 
The  two equations  then read
\be \la{KFEQ1} W-H + D_0(x-y) + \sum_{\a=1,2} \naa \maa \( ( \maa D_0
-1 ) \overline{ D_1^\aa }+
\overline{W} 
\, D_0 \) 
&=&0\, , \\
\la{KFEQ2} P+ D_0\(H-V' +W+n\overline{W}\) + \sum_{\a=1,2} \naa \(
\maa D_0 -1\) \overline{ D_1^\aa }&=&0\, .  \ee
The second equation involves an unknown linear function  
of $x$:
\begin{eqnarray}
 P(x, y) &\defeq& \frac1\beta\left\langle\text{tr}
 \frac{V'(x)-V'({\bf X})}{x-{\bf X}}{\bf H}(y)\right\rangle
    =  g_2H(y)+g_3 H_1(y) +x g_3H(y).  
\label{defP}
\end{eqnarray}

Equations \re{KFEQ1} and \re{KFEQ2} can be solved in favor of
$D_1^\ci$ or $D_1^\bu$.
If we define 
\begin{eqnarray}
 \AA^\ci &\defeq& \mci   D_0-1, \nonumber \\
 \BB^\ci &\defeq& (\mci -\mbu) \nci \{\mci  D_1^\ci
+ W \}
 -\mbu\(\overline{W}-\overline{V'} +H\)
 -x-y,  \label{defd1} \\
 \CC^\ci &\defeq&
 \mci  \mbu P+(\mci +\mbu)H-x+y
 -(\mci -\mbu)\(W+ \nci \overline{W}\)-\mbu V'\, ,
\no
\end{eqnarray}
and similarly for $\AA^\bu, \BB^\bu, \CC^\bu$, with an obvious
exchange $\ci\leftrightarrow \bu$, then \re{KFEQ1} and \re{KFEQ2} take
the following factorized form:
\begin{equation}
 \AA^\aa\overline{\BB^\aa} ~=~ \CC^\aa
 \qquad (\a=1,2) \, .
\label{D0D1}
\end{equation}
Equations \re{D0D1} are the main instrument of our analysis of the
\DJS boundary conditions.

  It is convenient to define the functions $D_0^\aa$ by
\begin{equation}
 D_0^\aa \defeq \frac{   D_0}{1-\maa  D_0}  \qquad (\a=1,2) \, .
  \la{DefDbw}
\end{equation}
  In Appendix \ref{sec:Derivation} we show that with this definition
  the recurrence equations (\ref{recestar}) hold also for $L=0$.  The
  equation for $L=0$ is a consequence of \re{D0D1}.

\subsection{Loop equation for the two-point function with Ord/Dir
boundary conditions}

 On the flat lattice, the \DJS boundary condition with $\nci=1$ is
 equivalent, for a special choice of the boundary parameters, to the
 Dirichlet boundary condition defined by the boundary factor
 \re{Dirichl}.  In order to make the comparison on the dynamical
 lattice, we will formulate and solve the loop equation for the
 two-point function of the BCC operator with ordinary/Dirichlet
 boundary conditions.
 
  Assume that the magnetic field points at the direction $a=1$.
   Then the  correlator in question is given in the matrix model by the
 expectation value
 \be
 \Omega( x,  y ) =  {1\over\b} \< \tr {\bf W}(x) {\bf R}(y)\>,
 \qquad {\bf R}(y) = {1\over y - {\bf X} -  B {\bf Y}_1}\, .
 \ee
 To obtain the loop equation we
  start with the identity
 \be
  W(x) - R(y)= (y-x) \O(x,y) - B \O_1(x,y),
 \ee
where we denoted by $R(y)$ the one-point function with Dirichlet
boundary condition,
\be
R(x) = {1\over\b} \< \tr {\bf R}(y)\>,
\ee
and introduced the auxiliary function
  \be \O_1(x,y) = {1\over\b} \< \tr { {\bf Y_1} \bf W}(x) {\bf R}(y)\>
  .  \ee 
The function $\O_1$ satisfies the identity
  \be \O_1(x,y ) =B\,  \oint_{{\cal C}_-} {dx_1\over 2\pi i} \ {\O(x_1, y
  )\O(-x_1, y ) \over x- x_1} \, , \ee
 which follows from \re{nonlocalWE}.  After symmetryzing with 
 respect to $x$ we get
   \be \O_1(x,y ) + \O_1(-x,y ) = B\,  \O(x,y) \O(-x,y ).  \ee
From here we obtain a quadratic functional equation for the correlator
$\O$:
 \be (x-y) \O(x, y ) -(x+y) \O(-x,y) + W(x) +W(-x) + B^2\,  \O (x,y)
 \O(-x,y)=2 R(y).  \ee
   The linear term can be eliminated by a shift
  \be
  G(x,y)= B\, \O(x,y ) -{x+ y\over B}\, .
\ee
The function $G$ satisfies 
\be
\la{eqG}   G(x,y)G(-x,y)
= -W(x)-W(-x) +2 R(y) - {x^2-y^2 \over B^2}  \, .
\ee
This equation is to be compared with the loop equation \re{D0D1} for
Ord/\DJS boundary conditions, with $\nci=1$ and $\mbu=0$:
\begin{equation}
  {1\over \mci} \AA^\ci(x,y) \BB^\ci (-x,y)~=~ -W(x) - W(-x) +H(y) -
  {x-y\over \mci} \, .
\label{D0D11}
\end{equation}
These two equations coincide in the limit $B, \mci\to\infty$.  To see
this it is sufficient to notice that in the limit $\mci\to\infty$ we
have the relation $\BB^\ci = y \AA^\ci$.  The exact relation between
$\AA^\ci$ (with $ \nci=1, \mbu=0$) and $G$ in this limit follows from
 the definitions of $D_0$
and $\O$:
  \be
   \AA^\ci(x, y')=
B  {G(x,y) - G(x, -y) \over 2y}  ,    
  \quad   \mci=B^2 ,\quad y'=y^2
  \qquad (B\to\infty).
  \ee
  %

%%%%%%%%%%%%%%%%%%%%%%%%%%%
\section{Scaling limit}
%%%%%%%%%%%%%%%%%%%%%%%%%%%
\label{sec:Scalimit}

In this section we will study the continuum limit of the solution, in
which the sum over lattices is dominated by those with diverging area
and boundary length.  The continuum limit is achieved when the
couplings $x$, $y$ and $\bar\mu$ are tuned close to their critical
values.
 
Once the bulk coupling constants are set to their critical values, we
will look for the critical line in the space of the boundary couplings
$y, \mci$ and $\mbu$.  After the shift \re{shiftx} the bondary
cosmological constant $x$ has its critical value at $x=0$, while
the critical value of $y$ in general depends on the values of $\mci$
and $\mbu$.

%%%%%%%%%%%%%%%%%%%%%%%%%%%%%%%% 
\subsection{Scaling limit of the disk one-point function}
%%%%%%%%%%%%%%%%%%%%%%%%%%%%%%%%% 
 
 Here we recall the derivation of the continuum limit of the one-point
 function $W(x)$ from the functional equation (\ref{funceqW}).  Even
 if the result is well known, we find useful to explain how it is
 obtained in order to set up the logic of our approach to the solution
 of the functional equations \re{D0D1}.

 In the limit $x\to 0$, the boundary length $|\p \Gamma|$ of the
 planar graphs in \re{defPPG} becomes critical.  The quadratic
 functional equation (\ref{funceqW}) becomes singular at $x\to 0$ when
 the coefficient $A$ on the r.h.s. vanishes.  This determines the
 critical value of the cosmological constant $\bar\mu$, for which the
 volume $|\Gamma|$ of a typical planar graph diverges.  The condition
 that the coefficient $B$ of the linear term vanishes determines the
 critical value of the temperature coupling $T=T_c$ for which the
 length of the loops diverges:
\begin{equation}
\la{critem} T_c = 1+\sqrt{\tfrac{2-n}{6+n}}\in[1,2] \, .
\end{equation}
Near the critical temperature the coefficient $B$ is proportional to
$T-T_c $.

We rescale $x\to \e x$, where $\e $ is a small cutoff parameter with
dimension of length, and define the renormalized coupling constants as
\begin{equation}\label{Ccouplings}
\bar\mu - \bar\mu_c \sim \e ^2\mu \,,\quad T-T_c \sim \e ^{2\th} t \,
\end{equation}
and write \re{WregWsin} as
\be \la{WregWsina} 
 W(x) \defeq W_\reg +  \e^{1+\th} w(x) \, .
 \ee
 The renormalized bulk and boundary cosmological constants are coupled
 respectively to the renormalized area $A$ and boundary length $\ell$
 of the graph $\G$ defined as
 \be
 A= \e^2\,  |\Gamma|, \qquad
 \ell = \e\,  |\p \Gamma|\, .
 \ee
 In the following we define the
 dimensions of the scaling observables by the way they scale with $x$.
 We will say that the quantity $f$ has dimension $d$ if the ratio
 $f/x^d$ is invariant with respect to rescalings.  In this case we
 write $[f]=d$.  The continuous quantities introduced until now have
 scaling dimensions
 \be [x]=1,\ [\mu] = 2, \ [t]= 2\th, \ [w] = 1+\th\, .  \ee

The scaling resolvent $w(x)$ is a function with a cut on the negative
axis in the $x$-plane.  It can be obtained from the general solution
found in \cite{EK} by taking the limit in which the cut extends to the
semi-infinite interval $[-\infty,-M]$.  To determine $M$ as a function
of $\mu$ and $t$ one has to solve a system of difficult transcendental
equations.  A simpler indirect method was given in \cite{Onthermal}.
We begin by noticing that the term $C x^4$ in (\ref{funceqW}) drops
out because it vanishes faster than the other terms when $x\to 0$,
and $B= B_1 \, t$, where $B_1$
depends only on $n$.  Introducing a hyperbolic map which resolves the
branch point at $x=-M$,
\begin{equation}\label{defxtau}
 x= M\cosh\tau\,,
\end{equation}
we obtain a quadratic functional equation for the entire function
$w(\tau)\equiv w[x(\tau)]$:
\begin{equation}\label{feqWtau}
 w^2(\tau+i\pi) +w^2(\tau)+ n\,w(\tau+i\pi)w(\tau) =
 A+B_1tM^2\cosh^2\tau\,.
\end{equation}
This equation does not depend on the cutoff $\e$, which justifies the
definition of the renormalized thermal coupling in \re{Ccouplings}.
Then the unique solution of this equation is, up a factor which
depends on the normalization of $t$,
\begin{equation}\label{xofct}
 w(\tau)=M^{1+\theta}\cosh(1+\theta)\tau
 +tM^{1-\theta}\cosh(1-\theta)\tau\,.
\end{equation}
One finds $B_1= 4\sin^2(\pi\theta)$ and
$A=\sin^2(\pi\theta)(M^{1+\theta}-tM^{1-\theta})^2$ for this solution.

The function $M=M(\mu, t)$ can be evaluated using the fact that the
derivative $\partial_\mu W(x)$ depends on $\mu$ and $t$ only through
$M$.  As a consequence, in the derivative of the solution in $\mu$ at
fixed $x$,
\be\la{dGc} \partial_\mu w= -M\partial_\mu M\left( (1+\theta) M^\theta
- (1-\theta)t M^{-\theta} \right) \frac{\sinh\theta\tau}{\sinh\tau},
\ee 
the factor in front of the hyperbolic function must be proportional to
$M^{\theta-1}$:
$$ 
 \partial_\mu M \Big( (1+\theta) M^\theta -(1-\theta) t M^{-\theta} \Big) 
 \sim  M^{\theta -1}.
$$
Integrating with respect to $\mu$ one finds, for certain normalization
of $\mu$,
\begin{equation}\label{bentr}
 \mu = (1+\theta)M^2 - t M^{2-2\theta}.
\end{equation}

To summarize, the disk bulk and the boundary one-point functions with
ordinary boundary condition, $-\p_\mu U$ and $-\p_x U$, are given in
the continuum limit in the following parametric form:
 \bigskip 
 \be\la{xoft} -\p_{x} U |_\mu &=& {M^{1+\th} } { \cosh((1+\th) \t )} \
 +   t \, {M^{1-\th } } {\cosh (1-\th )\t } , 
 \\ \no - \p_\mu U  |_{x}
 &\sim&\ \, \,   \, M^{ \th}\, \cosh \, \th \t ,\\ \no x\ \
 \ &=&\ \ \ M\, \cosh \t , \ee
with the function $M(\mu, t)$ determined from the transcendental
equation \re{bentr}.  The expression for $\p_\mu U$ was obtained by
integrating \re{dGc}.

The function $M(t,\mu)$ plays an important role in the solution.  Its
physical meaning can be revealed by taking the limit $x\to\infty$ of
the bulk one-point function $-\p_\mu U(x)$.  Since $x$ is coupled to
the length of the boundary, in the limit of large $x$ the boundary
shrinks and the result is the partition function of the $O(n)$ field
on a sphere with two punctures, the susceptibility $ u(\mu,t)$.
Expanding at $x\to\infty$ we find
 \be
  - \p_\mu U \sim x^{\th} - M^{2\th } \, x^{-\th} + \text{ lower  
 powers of } x \ee
 (the numerical coefficients are omitted).  We conclude that the
string  susceptibility is given, up to a normalization, by
 \be
 u  = M^{2\th}.
 \ee
The normalization of $u$ can be absorbed in the definition of the
string coupling constant $g_s\sim 1/\b$.  Thus the transcendental
equation \re{bentr} for $M$ gives the equation of state of the loop
gas on the sphere,
\be \la{stateeq} (1+\th)\, u^{1\over \th } - \, t\, u^{ 1-\th \over
\th} = \mu \,\, .  \ee

The equation of state \re{stateeq} has three singular points at
which the three-point function of the identity operator $\p_\mu u $
diverges.  The three points correspond to the critical phases of the
loop gas on the sphere.  At the critical point $t=0$ the
susceptibility scales as $u\sim \mu^\th$.  This is the dilute phase of
the loop gas, in which the loops are critical, but occupy an
insignificant part of the lattice volume.  The dense phase is reached
when $t/ x^\th \to -\infty$.  In the dense phase the loops remain
critical but occupy almost all the lattice and the susceptibility
has different scaling, $u\sim \mu^{1-\th\over\th}$.  The scaling of
the susceptibility in the dilute and in the dense phases match with
the values \re{cdilute} and \re{cdense} of the central charge of the
corresponding matter CFTs.  Considered on the interval $-\infty<t< 0$,
the equation of state \re{stateeq} describes the massless thermal flow
\cite{FSZ} relating the dilute and the dense phases.

At the third critical point $\partial_\mu M$ becomes singular but
$M$ itself remains finite.  It is given by
\be
t_c=\frac{1+\theta}{1-\theta}\, M_c^{2\theta} \ > 0,
\qquad
\mu_c =  - \th \frac{1+\theta}{1-\theta}\, M_c^{2 } \ < 0.
\ee
Around this critical point $ \mu-\mu_c\sim (M-M_c)^2 + \cdots $, hence
the scaling of the susceptibility is that of pure gravity, $u\sim
(\mu-\mu_c)^{1/2}$.

\subsection{The phase diagram  for  the \DJS boundary condition }

We found the scaling limit  of the one-point function \re{resW}
as a function of the renormalized bulk couplings, $\mu$ and $t$,
and the coupling $x$ characterizing the ordinary boundary.
Now, analyzing the loop equation \re{D0D1} for the two-point functions, 
we will look for the possible scaling limits for the couplings 
$y$, $\mci $ and $\mbu$, characterizing the \DJS boundary.

 As in the previous subsection, we will write down the conditions that
 the regular parts of the source terms ${\cal C}^\aa$ vanish.  Let us
 introduce the isotropic coupling $\l $ and the anisotropic coupling
 $\Delta$ as
\be \mci  = \l  +\hf \Delta,\ \mbu= \l -\hf \Delta\, \ee
and substitute \re{WregWsin} in the r.h.s. of (\ref{D0D1}).  We obtain
 \begin{eqnarray}
 \CC^\ci &=& c_0 + c_1 x + c_2x^2 - \, \Delta ( w+\nci \overline{w}
 )\,, \nonumber \\
   \CC^\bu &=& c_0 + c_1 x + c_2 x^2 + \, \Delta( w+\nbu
   \overline{w})\, ,
\label{D0D1scal}
\end{eqnarray}
where the coefficients $c_0$ and $c_1$ are functions of $\l $ and
$\Delta$:
\be c_0 &=& (\l ^2-{{\textstyle{1\over 4}}} \Delta^2) ( g_2 H+ g_3 H_1)+
2\l  H+y -\l  g_1 - \Delta\, {g_1 (\nci - \nbu) \over 2( 2+n)} \la{defmub}
\\
  c_1&=& (\l  ^2-{{\textstyle{1\over 4}}} \Delta^2) g_3 H -1- \l  g_2 +
  \Delta \, {g_2 (\nci-\nbu)\over 2(2-n)} \la{deftb}.  \\
c_2 &=& - g_3\( \l  + {\nci-\nbu \over 2(2+n)}\Delta\).
\ee

For generic values of the couplings $y, \l  $ and $ \Delta$, the
coefficient $c_0$ is non-vanishing.  The condition $c_0=0$ determines
the critical value $y_c $ where the length of the \DJS boundary
diverges.  \footnote{ Indeed, the term $c_0$ is the dominant term when
$x \to 0$.  For $c_0\ne 0$ the solution for $\AA^\aa$ and $\BB^\aa$ in
\re{D0D1} is given by linear functions of $ x$ and $ w$.  Such a
solution describes the situation when the length of the \DJS boundary
is small and the two-point function degenerates to a one-point
function.  } Once the boundary cosmological constant is tuned to its
critical value, the condition $c_1=0$ determines the critical lines in
the space of the couplings $\mci $ and $\mbu$, where the \DJS boundary
condition becomes conformal.  The two equations
\be\la{critconcc} c_0(y, \l  , \Delta)=0, \qquad c_1(y, \l  , \Delta)=0
\qquad (\mu=t=0) \ee
define a one-dimensional critical submanifold in the space of the
boundary couplings $\{ y, m , \Delta \}$:
\be \l  = \l ^*(\Delta)\, . 
 \la{curvaAS} 
\ee
Obviously $\AA^\ci$ and $\AA^\bu$ cannot be
simultaneously zero.  Therefore the curve \re{curvaAS} consists of two
branches, which correspond to different conformal \DJS boundary
conditions, the lines of anisotropic special transitions $AS_\ci$ and
$AS_\bu$:
\be
AS_\ci: & & \quad \AA^\ci = 0, \ \AA^\bu\ne0;\no\\
AS_\bu: & & \quad \AA^\bu = 0, \ \AA^\ci\ne0.
\ee

The branch $AS_\ci$ corresponds to $\Delta>0$,
while the branch $AS_\bu$ corresponds to $\Delta<0$.  Consider the
behavior of correlators $D_0^\aa$ on the two branches of the critical
line.  By the definition \re{DefDbw}, the two correlators $D_0^\ci$
and $D^\bu_0$ are related by
\begin{equation}\label{relD0}
D_0^\ci =\dfrac{D_0^\bu}{1- \Delta D_0^\bu},\quad
D_0^\bu=\dfrac{D_0^\ci}{1+ \Delta D_0^\ci}.
\end{equation}
On the branch $AS_\ci$ the correlator $D^\ci_0$ diverges while $D^\bu_0$
remines finite, and {\it vice versa}.  Assume that $\mci$ and $\mbu$
are positive.  Then $D^\aa_0$ are both positive by construction.  If
$\Delta=\mci -\mbu>0$, then the coefficients of the geometric series
\begin{equation}\label{expD0}
{D_0^\ci }=\sum_{k=0}^\infty{\Delta^k \(D_0^{\bu}\)^{k+1}}
\end{equation}
are all positive and $D^\ci_0$ diverges, while $D^\bu_0\to 1/\Delta$.
Thus the branch $AS_\ci$, where $D_0^\ci$ diverges, is associated with
$\Delta>0$.  On this branch the probability that the loops of color
$\ci$ touch the \DJS boundary is critically enhanced.  In the
correlator $D_0^\ci $, the ordinary boundary behaves as a loop of type
$\ci$ and can touch the \DJS boundary.  The geometrical progression
\re{expD0} reflects the possibility of any number of such events, each
contributing a factor $\Delta$.  Conversely, the ordinary boundary for
the correlator $D^\bu_0$ behaves as a loop of type $\bu$, since such
loops almost never touch the \DJS boundary.  On the branch $AS_\bu$
the situation is reversed.

It is not possible to solve explicitly the conditions of criticality
\re{critconcc} without extra information, because they contain two
unknown functions of the three couplings, $H$ and $H_1$.
Nevertheless, the qualitative picture can be reconstructed.

First let us notice that the form of the critical curve can be
evaluated in the particular cases $\nci=n$ and $\nci=0$.  In the first
case $\nbu=0$ and the correlation functions do not depend on $\mbu$
and so the coefficients $c_1$ and $c_2$ depend on $\l $ and $\Delta$
through the combination $\mci = \l  + \Delta/2$.  Similarly one
considers the case $\nci=0$.  The phase diagram in these two cases
represents an infinite straight line separating the ordinary and the
extraordinary transitions:
   \be\la{crcrv}
   \l ^*(\Delta) = 
   \begin{cases}
         \l  _c - \Delta/2     & \text{ if } \nci = n, \\
      \l _c + \Delta/2   & \text{if } \nci = 0.
\end{cases}
    \ee
The critical line crosses the axis $\Delta=0$ at the special point
$\l =\l _c$.  The value of $\l _c$ can be evaluated by solving \re{critconcc}
for $\nci=n$ and $\mbu=0$.  The result is
\be\la{defmsp} 
\l _c   = {\sqrt{(6+n)( 2-n)}\over 1-n} .
 \ee
 For general $\nci\in [0, n]$ we can determine three points of the
 critical curve:
 \be\la{specialpts} \{ \mci , \mbu\} =\{ {1-n\over 1-\nci}\l _c , \ 0 \}
 ,\quad \{ 0,\ {1-n\over 1-\nbu}\l _c \}, \quad \{\l _c ,\l _c \}\, .  \ee

\begin{figure}
\begin{center}
\includegraphics[width=200pt]{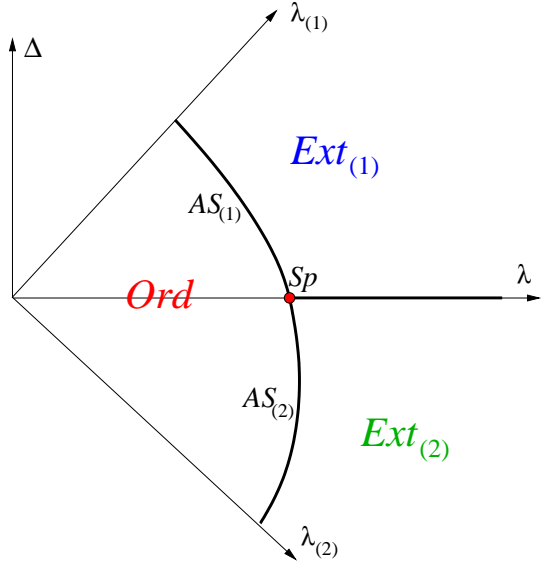}
\end{center}
\caption{\small Phase diagram in the
rotated $({\mci}, {\mbu})$ plane  for $n<1$ and $\nci>\nbu$.
The ordinary and the extraordinary phases are separated by a
 line of anisotropic special transitions, which  
consists of two branches, $AS_\ci$ and $AS_\bu$.
The two extraordinary phases, $Ext_\ci$ and $Ext_\bu$,
 are separated by the isotropic line $\Delta=0$}

\label{fig:FhaseDR}
\end{figure}

In the two limiting cases considered above the critical line, given by
equation \re{crcrv}, crosses the anisotropic line without forming a
cusp.  Is this the case in general?  Let us consider the vicinity of
the special point $(\l ,\Delta)= (\l _c, 0) $.  In the vicinity of the
special point a new scaling behavior occurs.  In this regime the term
$x^2$ in \re{D0D1scal} cannot be neglected.  The requirement that all
terms in \re{D0D1scal} have the same dimension determines the scaling
of the anisotropic coupling $\Delta$:
\be\la{scalD}
 [\Delta]= 1-\th.
\ee
In order to determine the form of the critical curve near the special
point, we return to the equations \re{critconcc} and consider the
behavior near the special point of the unknown functions $H(y)=
{1\over\b} \< \tr {\bf H}(y) \>$ and $H_1(y)= {1\over\b} \< \tr {\bf
X}{\bf H}(y) \>$, which depend implicitly on $\l $  and $\Delta$.  For
$t=\mu=0$, these functions can be decomposed, just as the one-point
function with ordinary boundary condition, $W(x)$, into regular and a
singular parts:
\be \la{HregWsin} H(y) = H^\reg(y) + h(y), \quad H_1(y) = H_1^\reg(y)+
\text{cst} \cdot h(y) \, .  \ee 
On the critical curve $\l = \l ^*(\Delta)$
the singular part of $H$ vanishes and the coefficient $c_1$ given by
eq.  \re{deftb} can be Taylor expanded in $\l -\l _c $ and $\Delta$:
\be
c_1(m, \Delta) \equiv
A_1 \, (\l -\l _c ) 
+   B_1\, (\nci-\nbu)  \Delta
+A_2 \, (\l -\l _c )^2 
+ B_2  \, \Delta^2 +\dots = 0.
\ee
If $A_1\ne 0$, the critical curve is given by a regular function of
$\l $  and $\Delta$, the critical curve is a continuous line which
crosses the real axis at $\l =\l _c $ without forming a cusp.  This form
of the curve differs from the predictions of \cite{DiEi} and
\cite{DJHS2}, where a cusp-like form is predicted by a scaling
argument.  We will see later that the fact that the critical curve is analytic 
at the special point  does not contradict the scaling \re{scalD}.

%%%%%%%%%%%%%%%%%%%%%%%%%%%%%%%%%%%% 
\subsection{Scaling limit of the functional equation for the disk
two-point function}
%%%%%%%%%%%%%%%%%%%%%%%%%%%%%%%%%%%% 

 \subsubsection{The scaling limit for $\Delta\ne0$}

Consider first the case when the anisotropic coupling $\Delta$ is
finite and assume that we are on the branch $AS_\ci$ where $\Delta>0$.
Then the $x^2$ term on the r.h.s. can be neglected, because it is
subdominant compared to $w\sim x^{1+\th}$.  The scaling limit
corresponds to the vicinity of the critical submanifold where the two
coefficients $ c_0$ and $c_1$ scale respectively as $ x^{ 1+\th}$ and
$ x^{ \th}$.

We are now going to find the scaling limit of the loop equations
\re{D0D1}.  In the scaling limit we can retain only the singular parts
of the correlators $D_L^{\ci, \bu}$, which we denote by $d_L^{\ci,
\bu}$ ($L=0,1,...$).  We define the functions $d_0^\ci$ and $d_0^\bu$
by
\be D_0^\ci = \, d_0^\ci,\quad D_0^\bu = {1\over \Delta - \, d_0^\bu}
\approx {1\over\Delta} + {1\over\Delta^2} \, d_0^\bu .  \ee
Then the relation  \re{relD0} implies
\be
d_0^{\ci } \, d_0^{ \bu} = -1.
\la{dode}
\ee
We define in general $d_L^{\ci, \bu}$ as the singular part of
$D_L^{\ci, \bu}$, with the normalization chosen so that the recurrence
equation \re{recestar} holds for any $L\ge 0$:
\be\la{recdL} d^\ci_{L+1} &=& w\star d^\ci_L\, , \no\\
 d^\bu_{L+1} &=& w\star d^\bu_L  .
  \ee
On the branch $AS_\ci$ the first of the two equations \re{D0D1}
becomes singular, since $\AA^\ci$ vanishes while $\AA^\bu$ remains
finite.  We write this equation in terms of $d_0^\bu$ and $d_1^\ci$
using that
 \be \AA^{\ci}= \mci  \, d_0^\bu , \quad \BB^{\ci}= {\Delta \over
 \mci } \, d_{1}^{\ci} .  \la{defdd} \ee
  We get
 \be
  \hskip 3cm
  { d^\bu _0} \, \overline{d^\ci_1} + w +\nci \overline{w} =
 \mu_B-t_Bx \, \, , 
 \hskip 3cm
 (\Delta>0)\la{scdd1} \ee
where $\mu_B$ and $t_B$ are defined by
 \be\la{defmuB} {c_0\over\Delta}= \mu_B, \qquad {c_1\over\Delta}= -
 t_B\, .  \ee
Once the solution of \re{defmuB} is known, all two-point functions
$d_L^\aa$ can be computed by using the recurrence equations \re{recdL}.

The scaling limit near the branch $AS_\bu$ ($\Delta<0$) is obtained by
using the symmetry $\nci\leftrightarrow \nbu, \ \Delta \leftrightarrow
-\Delta$.  In this case one obtains another equation
 \be
  \hskip 3cm
  { d^\ci _0} \, \overline{d^\bu_1} + w +\nbu \overline{w} =
 \mu_B-t_Bx \, \, .
 \hskip 3cm
 (\Delta<0)\la{scdd2} \ee
Note that the relation \re{dode} is true on both branches of the
critical line.

The map $\{ y, \l  \} \to \{ \mu_B, t_B\}$ defined by \re{defmub},
\re{deftb} and \re{defmuB} represents a coordinate change in the space
of couplings which diagonalizes the scaling transformation.  The
coupling $\mu_B $ is the renormalized boundary cosmological constant
for the \DJS boundary.\footnote{More precisely, it is a combination of
the boundary coupling constant and the disk one-point function with
\DJS boundary.  What is important for us is that the condition
$\mu_B=0$ fixes the critical value of the bare \DJS cosmological
constant $y$.  At $\mu_B=0$, the length of the \DJS boundary diverges.
} The coupling $t_B$
is the renormalized boundary matter coupling, which defines the \DJS 
boundary condition.  The dimensions of these couplings are
\be
[\mu_B]= 1+\th, \qquad
[t_B]= \th.
\ee
Once we choose $y$ so that $\mu_B=0$, the condition $t_B=0$ gives the
critical curve where the anisotropic special transitions take place.
If the function $t_B$ is regular near the critical line
$\l =\l ^*(\Delta)$, then it can be replaced by the linear approximation
 \be t_B \sim \l ^*(\Delta)- \l  .  \la{lintB}\ee

 The deformations in the directions $t_B$ and $\Delta$, are driven by
 some Liouville dressed boundary operators $\CO_{t_B}^B$ and
 $\CO_{\Delta}^B$.  Knowing the dimensions of $t_B $ and $\Delta$, we can
 determine the Kac labels of these operators with the help
 of the KPZ formula.  The general rule for evaluating the Kac labels
 in 2D gravity with matter central charge \re{cdilute} is the
 following.  If a coupling constant has dimension $\a$, then the
 corresponding operator has Kac labels $(r,s)$ determined by
   \be \a =\a_{r,s}= \text{min} \( 1+{\th \over 2} \pm { (1+\th) r - s
   \over 2}\).  \ee
The details of the identification are given in Appendix
\ref{sec:Liouv}.  We find
 \be
\CO_{t_B}^B=  {\cal O}_{1,3}^B,
\qquad
\CO_{\Delta}^B ={\cal O}_{3,3}^B.
\ee
Near the special point we have 
 \be\la{sctB}
 \Delta\sim t_B^{1/\phi}, \qquad
 \phi = {\th\over 1-\th}
  = {\a_{1,3}\over\a_{3,3}}
<1.
 \ee
Since $t_B$ and $\l $ scale differently, there is no contradiction between the
scaling \re{sctB} and the analyticity of
the critical curve near the special point.

\subsubsection{The scaling limit in  the isotropic direction $(\Delta =0)$}

Along the isotropic line $\Delta = 0$ the two functional equations
\re{D0D1} degenerate into a single equation for the correlator $D_0$:
\be \CA \equiv \l  D_0-1 = - { y-x + \l (2H+ \l  P-V') \over y - x +  \l \(W+H -
V'\)} .  \ee
In order to evaluate $ D_1= D_1 ^\ci= D_1^\bu $, we can consider the
linear order in $\Delta$.  It is however easier to use the fact that
$D_0$ and $D_1$ do not depend on the splitting $n=\nci+\nbu$.
Furthermore, if we choose $\nci=n$ and $ \mci =\l  $, the observables do
not depend on $\mbu$, which can be chosen to be zero.  Taking
$\nci=n$, $\mci =m$ and $ \nbu=\mbu=0$, we obtain from \re{D0D1}
\be\la{hatle} \overline{ \BB^{\ci}} \big|_{\nci=n } \equiv  \l \( \l 
\overline{D_1} + n \overline{W }\)+x-y = {y-x +  \l (H-W- n \overline{W}
)\over  \l  D_0-1}\, .  \la{Bsp}\ee

From these expressions it is clear how the scaling of the singular
parts of $D_0$ and $D_1$, which we denote respectively by $d_0 $ and
$d_1$, change when we go from $ \l =0$ to $\l =\l _c $.  When $ \l =0$ we have
$H(y)=W(y)$ and $D_0$ is the disk partition function with ordinary
boundary conditions and two marked points on it, eq.  \re{DNNo}.  When
$\l =\l _c $ and $y=y_c $,
\be d_0 = g_3{x^2\over w} \sim x^{1-\th}, \qquad d_1 = { w(w+n\bar
w)\over g_3\ x^2 }\sim x^{2\th} \qquad (\l =\l _c , \ y=y_c ).  \ee

\subsubsection{ Dirichlet versus \DJS}

%%%%%%%%%%%%%%%%%%%%%%%%%%%%%%%%%

Now we will focus on the special case $\nci=1$ and compare the scaling
behavior with that for the Dirichlet boundary conditions.  The
critical behavior of the two-point correlator in both cases is the
same, but the boundary coupling constants correspond to different
boundary operators.

Consider the functional equation \re{D0D11} for the correlator
with Ord/\DJS boundary conditions when $\nci=1$.  The critical value
of $\mci$ is infinite in this case, see equation \re{specialpts}.  The
scaling limit of \re{D0D11} is
 \be d_0^\ci(x,y) d_1^\bu (-x,y)~=~ -w(x) - w(-x) + \mu_B - {x \over
 \mci} \, .
\label{D0sc}
\ee
The last term remains finite if $\mci$ tends to infinity as
$x^{-\th}$.  The scaling boundary coupling can be identified as $t_B=
1/\mci$ and equation \re{D0sc} takes the general form \re{scdd1}.
What is remarkable here is that the boundary temperature constant need
not to be tuned.  Equation \re{D0sc} holds for any value of $\mci$.
On the other hand, when $t_B= 1/\mci$ is small, the last term
describes the perturbation of the $AS_\ci$ boundary condition by the
thermal operator $\CO_{1,3}$ with $\a_{1,3}=\th$.  When $\mci$ is
small, the last term accounts for the perturbation of the ordinary
boundary condition by the two-leg boundary operator $\CO_{3,1}$
with $\a_{3,1}= -\th$, whose matter component is an is irrelevant
operator.

Now let us take the scaling limit of the quadratic functional equation
\re{eqG} for the correlator with Ord/Dir boundary conditions.  At
$x=0$ the equation \re{eqG} becomes algebraic.  The critical value $y=
y_c$, where the solution develops a square root singularity, is
determined by
$$
2W(0) - 2R(y_c) +{y_c^2 / B^2 } =0.
$$
We can write   equation \re{eqG} as
\be \la{eqGsl} G(x,y)G(-x,y) = \mu_B -w(x)-w(-x) - {x^2 \over B^2 } \,
, \ee
where
\be \mu_B =2 w(0) +2 [R(y)- R(y_c)] +{y^2- y_c^2 \over B^2 }.  \ee
For any finite value of $B$, the scaling limit of this equation 
is 
\be \la{eqGsc} G(x,y)G(-x,y) = \mu_B -w(x)-w(-x) \, .  \ee
The $x^2$ term survives only if  $B$   vanishes as $x^{1-\th\over 2}$:
\be
[B] =  (1-\th)/2 = \a_{2,1} .
\ee
This is the expected answer, because $\CO_{2,1}^B$ is the one-leg
boundary operator which creates an open line starting at the boundary.
We conclude that the Dirichlet and the \DJS boundary conditions have
the same scaling limit, but in the first case the boundary coupling
$\l $  corresponds to a relevant perturbation and it is sufficient give
it any finite value, while in the second case the boundary coupling
$\mci$ corresponds to an irrelevant perturbation and therefore must be
infinitely strong.

%%%%%%%%%%%%%%%%%%%%%%%%%%%%%%%%%%%% 
\section{Spectrum of the  boundary  operators}
%%%%%%%%%%%%%%%%%%%%%%%%%%%%%%%%%%%%%
\label{sec:Spectr}

Let us denote by $\alpha^\ci_L$ and $\alpha^\bu_L$ the scaling
dimensions respectively of $d^\ci_L$ and $d^\bu_L$:
\be
\alpha^\aa_L =[d^\aa_L] 
\qquad (L\ge 0, \ a=1,2).
\ee
The recurrence equations \re{recdL} tell us that the dimensions grow
linearly with $L$:
\be\la{additiv}
\a_L^\aa =L  [w] + \a_0^\aa \, .
\ee
 These relations make sense in the dilute phase, where $[w]= 1+\th$,
 as well as in the dense phase, where $[w]= 1-\th$.  In addition, by
 \re{dode} we have
\be
\a_0^\ci +\a_0^\bu=0.
\ee
Thus all scaling dimensions are expressed in terms of $\a_0^\ci$.

Let us evaluate $\a_0$ for the branch $AS_\ci$ of the critical line.
We thus assume that $\Delta$ is finite and positive sufficiently far
from the isotropic special point.  Take $ \mu= \mu_B=t_B=0$ and write
the shift equations which follow from \re{scdd1},
  \be\la{eqexshift} AS_\ci: \qquad {d^\ci_0(e^{i\pi} x)\over
  d^\ci_0(e^{-i\pi} x)} = {w (e^{-i\pi} x ) +\nci w ( x) \over w
  (e^{i\pi} x ) +\nci w ( x) } .  \ee
The one-point function \re{xofct} behaves as $w\sim x^{1+\th} $ in the
dilute phase ($t=0$) and as $w\sim x^{1-\th} $ in the dense phase
($t\to-\infty$).  In both cases the r.h.s. is just a phase factor.
Since all the couplings except for $x$ have been turned off,
$d_0^\ci(x)$ should be a simple power function of $x$.
Substituting $d_L^\ci(x)\sim x^{\a_L^\ci}$ in
\re{eqexshift}, we find
  \be\la{eqnaa} 
  \nci= {\sin\pi (\a_0^\ci \pm \th)\over \sin
  \pi\a_0^\ci}\, ,  \qquad \(+ \ \text{for dilute}, \ - \ \text{for
  dense}\)\,.  \ee
This equation determines the exponent $\alpha^\ci_0$ up to an integer.
In the parametrization \re{paramen} we have $\a_0^\ci= - \th r
+j_{\text{dil}}$ in the dilute phase and $\a_0^\ci= \th r +
j_{\text{den}}$ in the dense phase.
  
The integers $j_{\text{dil}}$ and $j_{\text{den}}$ can be fixed by
additional restrictions on the exponents.  Let us assume that $ \mci
 $ and $ \mbu $ are non-negative, $n\ge 0$ and the boundary
parameter $r $ is in the `physical' interval $1\le r \le 1/\th - 1$,
where both $\nci$ and $\nbu$ are non-negative.  These assumptions
guarantee that the Boltzmann weights are positive and the loop
expansion of the observables has good statistical meaning.  Since all
loop configurations that enter in the loop expansion of the one-point
function $W(x)$ are present in the loop expansions of $ \AA^{\aa} $
and $\BB^{\aa} $, the singularity of these observables when $\l \to
\l ^*  (\Delta)$ must not be weaker than that of $W$.  In other words, the
scaling dimensions of $ d_0^\bu \sim \AA^{\ci}$ and $d_1^\ci\sim
\BB^{\ci}$ must not be larger than the scaling dimension of the
one-point function $w$:
 \be \la{boundsAB}
 \a_0^\bu<[w], \quad \a_1^\ci<[w]
 .
  \ee
Since $ \a_0^\bu+\a_1^\ci= [w]$, this also means that $ \a_0^\bu$
and $\a_1^\ci$ are non-negative.  
Taking into account that $[w]= 1\pm\th$ in the dilute/dense phase, 
we get  the bound
 \be - (1\pm\th) \le \alpha^{\ci}_0 \le 0 \qquad \(+ \ \text{for
 dilute},  \ - \ \text{for dense}\) . \ee
This bound determines $j_{\text{dil}}=0$ and $ j_{\text{den}}=-1$.  As
a consequence, on the branch $AS_\ci$ of the critical line the
dimensions $\alpha^\ci_L =[d^\ci_L]$ in the dilute and in the dense
phases are given by
\begin{equation}\label{scalingd}
 AS_\ci:\quad \begin{array}{lll} \alpha^\ci_L = L(1+\theta) -\theta r
 ,& \a_L^\bu = L(1+\th) + \th r & \qquad \text{(dilute phase)} \\
 \alpha^\ci_L = L(1-\theta) +\theta r  -1, & \a_L^\bu=
 L(1-\th) - \th r + 1 & \qquad \text{(dense phase)} .
\end{array}
\end{equation}
 Note that the results for the dense phase are valid not only in the
 vicinity of the critical line $AS_\ci$, but in the whole half-plane
 $\Delta>0$.
  
By the symmetry $\ci\leftrightarrow \bu$, the exponents $\alpha^\ci_L$
on the branch $AS_\ci$ and the exponents $\alpha^\bu_L$ on the branch
$AS_\bu$ should be related by $\nci\leftrightarrow \nbu$, or
equivalently $r \leftrightarrow 1/\theta-r $:

\begin{equation}\label{scalingdba}
 AS_\bu:\quad\begin{array}{lll} \alpha^\ci_L = L(1+\theta) -\theta r 
 +1 ,& \alpha^\bu_L = L(1+\theta) +\theta r  -1 , & \qquad
 \text{(dilute phase)} \\
 \alpha^\ci_L = L(1-\theta) +\theta r , & \alpha^\bu_L = L(1-\theta)
 -\theta r &\qquad \text{(dense phase)} .
\end{array}
\end{equation}

The scaling exponents of the two-point functions of the $O(n)$ model
coupled to 2D gravity allow, through the KPZ formula \cite{KPZ, DDK},
to determine the conformal weights of the matter boundary operators.
In the dilute phase, where the Kac parametrization is given by
\re{KacTdil}, the correspondence between the scaling dimension $\a$ of
a boundary two-point correlator and the conformal weight $h _{r,s}$ of
the corresponding matter boundary field is given by
\begin{equation}\la{KPZdil}
\alpha = (1+\theta)r-s \quad \to\quad h = h _{r,s}\, \qquad
\text{(dilute phase).}\
\end{equation}
In the dense phase, where the Kac labels are defined by \re{KacTden},
one obtains, taking into account that the identity boundary operator
for the ordinary boundary condition has `wrong' dressing,
\begin{equation}\la{KPZden}
\alpha = r - s(1-\theta) \quad\to\quad h = h _{r,s}\, \qquad
\text{(dense phase)}.
\end{equation}
From \re{KPZdil} and \re{KPZden} we determine the scaling dimensions
of the $L$-leg boundary operators \re{SnJS}:
\begin{equation}\label{scalingdci}
AS_\ci:
\quad
\begin{array}{lll}
S_L^\ci \to \CO^B_{r-L, r}, & S_L^\bu \to \CO^B_{r+L, r} &
\text{(dilute phase)}\\
 S_L^\ci \to \CO^B_{r-1, r -L}, 
 &
   S_L^\bu \to \CO^B_{r-1, r+L} \qquad
 &
 \text{(dense phase)}. 
\end{array}
\end{equation}
\begin{equation}\label{scalingdbu}
AS_\bu:
\qquad
\begin{array}{lll}
S_L^\ci \to \CO^B_{r-L, r+1}, 
& 
S_L^\bu \to \CO^B_{r+L, r+1}
&\qquad  \text{(dilute phase)}\\
 S_L^\ci \to \CO^B_{r, r -L},  &  
  S_L^\bu \to \CO^B_{r, r+L}  
 &
\qquad  \text{(dense phase)}. 
\end{array}
\end{equation}
These conformal weights are in accord with the results of
\cite{Jacobsen:2006bn}, \cite{Kostov:2007jj}, \cite{Bourgine:2008pg},
\cite{DJHS2}.  We remind that the scaling dimensions are determined up
to a symmetry of the Kac parametrization:
\be h_{r,s} = h_{-r,-s},\quad h_{r, s-1} = h_{r+ 1/\th, s+1/\th}
\qquad \text{(dilute phase)}\\
h_{r,s} = h_{-r,-s}, \quad h_{r+1,s} = h_{r+ 1/\th, s+1/\th} \qquad
\text{(dense phase)} .  \ee

Comparing the scaling dimensions in the dilute and in the dense phase,
we see that the bulk thermal flow $t \, \CO_{1,3}$ transforms the
boundary operator $\CO_{r, s} $ in the dilute phase into the boundary
operator $\CO_{s-1, r}$ in the dense phase.  For the rational points
$\th = 1/p$, our results for the endpoints of the bulk thermal flow
driven by the operator $t \, \CO_{1,3}$ match the perturbative
calculations performed recently in \cite{Fredenhagen:2009sw}.

In the $O(n)$ model the boundary parameter $r$ is continuous and we
can explore the limit $r\to 1$,  in which the BCC
operator $\CO_{r,r}$ 
carries  the same Kac labels as the identity operator. Let us call this operator
 $\tilde \CO_{1,1}^B$.
The bulk thermal flow  transforms the operators $\tilde \CO_{1,1}^B$ and 
$\CO_{1,1}^B$ into two different boundary operators in the dense phase.
Hence there are at least two distinct boundary operators 
with Kac labels $(1,1)$: the identity operator and the  limit $r\to 1$
of the operator  $ \CO_{r,r}^B$.

   %%%%%%%%%%%%%%%%%%%%%%%%%%%%%
\section{Solution of the loop equations in the scaling limit}
%%%%%%%%%%%%%%%%%%%%%%%%%%%%%
 \label{sec:Solution}

We are going to study two particular cases where the analytic solution
of the functional equations \re{scdd1} and \re{scdd2} is accessible.
First we will evaluate the two-point function on the two branches
$AS_\ci$ and $AS_\bu$ of the critical line, where the $O(n)$ field is
conformal invariant both in the bulk and on the boundary.  In this
case $t=t_B=0$ and the boundary two-point function is that of
Liouville gravity.  The three couplings are introduced by the world
sheet action of Liouville gravity with matter central charge
\re{cdilute}, which we write symbolically as
 \be {\cal S}_{\text {Liouv}} = {\cal S}_{\text {free}} + \int \limits
 _{\text{bulk}} \mu\, \CO_{1,1} + \hskip -0.6cm \int \limits
 _{\text{Ord.  boundary}}\hskip -0.6cm x\, \CO_{1,1}^B+\hskip
 -0.5cm\int \limits _{\text{\DJS boundary}} \hskip -0.5cm \mu_B\,
 \CO_{1,1}^B .  \la{ttBAL} \ee
Since the perturbing operators in this case are Liouville primary
fields,
$$
\CO_{1,1}\sim e^{2 b\phi}, \quad \CO_{1,1}^B\sim e^{b\phi},
$$
the two-point function is given by the product of matter and Liouville
two-point functions.  Up to a numerical factor, the solution as a
function of $\mu$ and $\mu_B$ must be given by the boundary two-point
function in Liouville theory \cite{FZZb}.  We will see that indeed the
functional equation \re{scdd1} is identical to a functional equation
obtained in \cite{FZZb} using the operator product expansion in
boundary Liouville theory.
 
In the second case we are able to solve, we take $\mu=\mu_B=0$ and
non-zero matter couplings $t$ and $t_B$.  This case is more
interesting, because it is not described by the standard Liouville
gravity.  The corresponding world sheet action is symbollically
written as
\be {\cal S} = {\cal S}_{\text {free}} + \int \limits_{\text{bulk}}
t\, \CO_{1,3} + \hskip -0.6cm\int\limits _{\text{Ord.
boundary}}\hskip -0.5cm x\, \CO_{1,1}^B+ \hskip -0.6cm\int
\limits_{\text{\DJS boundary}}\hskip -0.5cm t_B\, \CO_{1,3}^B .
\la{ttBA} \ee
The worldsheet theory described by this action is more complicated
than Liouville gravity, because it does not enjoy the factorization
properties of the latter.  The boundary two-point correlator does not
factorize, for finite $t$ and/or $t_B$, into a product of matter and
Liouville correlators, as is the case for the action \re{ttBAL}.  This
is because the perturbing operators $\CO_{1,3}$ and $\CO_{1,3}^B$ have
both matter and Liouville components: $$ \CO_{1,3} \sim \Phi_{1,3}\,
e^{2b(1-\th)\phi}, \quad \CO_{1,3}^B\sim \Phi_{1,3}^B\,
e^{b(1-\th)\phi}.
 $$
Let us mention that the theory of random surfaces described by the
action \re{ttBA} has no obvious direct microscopic realization.  Our
solution interpolates between the two-point functions for the dilute
($t=0$) and the dense ($t\to-\infty$) phases of the loop gas, on one
hand, and between the anisotropic special $(t_B=0)$ and
ordinary/extraordinary boundary conditions ($t_B\to\pm\infty$), on the
other hand.

\subsection{Solution for  $t=0, t_B=0$ }

In the dilute phase ($t=0$) the solution \re{defxtau} - \re{xofct} for
the boundary one-point function takes the form
\be x= M\cosh\t, \quad w(x)= M^{1+\th} \cosh(1+\th)\t.  \ee
Then the loop equations \re{scdd1} become a shift equation
\be  d^\bu_0(\t) \, d^\ci_1(\tau\pm i\pi) +
w_0 \cosh\[(1+\theta)\tau \pm i \pi (1-r) \th \]
 =  \mu_B-t_B M\cosh\tau  \, ,  
 \la{scddtau} \ee
where we  introduced the constant
\be w_0 = M^{1+\theta} {\sin \pi\th\over \sin\pi r\th} .  \ee

At the point $t_B=0$, where the \DJS boundary condition is conformal,
this equation can be solved explicitly.  After a shift $\t\to\t \mp
i\pi$ we write it, using \re{dode}, as
\be\, d^\ci_1(\tau) = \[ w_0 \cosh\[(1+\theta)\tau \pm i \pi r \th \]
- \mu_B\, \]\ d^\ci_0(\t \pm i\pi) .  \la{scddtauc} \ee

 If we parametrize $\mu_B$ in terms of a new variable $\sigma$ as
\begin{equation}\la{paramubdil}
 \mu_B ~=~ w_0 \cosh(1+\theta)\sigma,
\end{equation}
the loop equation turns out to be identical to the functional identity
for the boundary Liouville two-point function \cite{FZZb}, which we
recall in Appendix \ref{sec:Liouv}.  In the Liouville gravity
framework, $\tau$ and $\sigma$ parametrize the FZZT branes
corresponding to the ordinary and anisotropic special boundary
conditions.

The loop equations for the dense phase ($t\to-\infty$), are given by
\re{scddtauc} with $\theta$ sign-flipped.  This equation describes the
only scaling limit in the dense phase.  The term with $t_B$ is absent
in the dense phase, because it has dimension $1+\th$, while the other
terms have dimension $1-\th$.  In this case, the loop equation gets
identical to the functional identity for the Liouville boundary
two-point function if we parametrize $\mu_B$ as
\begin{equation}
 \mu_B ~=~ w_0 \cosh\big(1-\theta)\sigma.  \la{paramubden}
\end{equation}

%%%%%%%%%%%%%%%%%%%%%%%%%%%%%%%%%%%%% 
\subsection{ Solution for $\mu =0, \ \mu_B=0$ }
%%%%%%%%%%%%%%%%%%%%%%%%%%%%%%%%%%%%%% 

Here we solve the loop equation \re{scdd1} in the scaling limit with
$\mu = \mu_B=0$ but keeping $x$, $t$ and $t_B$ finite.  Let us first
find the expression for the one-point function $w(x)$ for $\mu=0$.
The equation \re{bentr} has in this case two solutions, $M=0$ and $M=
( 1+\th)^{-1} t^{1\over 2\th}$.  One can see \cite{Onthermal} that the
first solution is valid for $t<0$, while the second one is valid for
$t>0$.  Therefore when $\mu =0$ and $t\le 0$, the solution
\re{defxtau}-\re{xofct} takes the following simple form:
\be w(x)= x^{1+\theta}+tx^{1-\theta} 
\qquad (t\le 0).  \ee

Introduce  the following exponential parametrization 
of $x, t, t_B$ in terms of
$\t, \g,\tilde\g$:
\be x=e^\tau, \quad t=-e^{2\gamma\theta}, \quad t_B =-2w_0 \,
e^{\gamma\theta}\sinh(\tilde\gamma\theta),\quad w_0 = \frac{
\sin(\pi\theta)}{\sin(\pi r  \theta)}.\ee
 In terms of the new variables, equation \re{scdd1} with $\mu_B=\mu=0$
 acquires the form
\be { d^\ci_1(\tau)/ d^\ci_0(\t \pm i\pi)} &=& w_0 \( -2
e^{\gamma\theta}\sinh(\tilde\gamma\theta) e^\t + e^{(1+\theta)\tau \pm
i \pi r \th} -e^{ 2\th \g } e^{(1+\theta)\tau \pm i \pi r \th}\) \,
\no \\
 &=& 4 w_0 \, e^{\tau+\gamma\theta}
 \cosh\frac{\theta(\tau-\gamma+\tilde\gamma\pm i\pi r  )}2
 \sinh\frac{\theta(\tau-\gamma-\tilde\gamma\pm i\pi r  )}2\,.
 \qquad
 \label{lpscl}
\ee
Taking the logarithm of both sides we obtain a linear difference
equation, which can be solved explicitly.  The solution is given by
\begin{equation}
AS_\ci:
\quad
\begin{array}{lll}
d_0 ^\ci(\tau)&=& {1\over w_0} e^{-\frac\tau2-\gamma( r 
\theta-\frac12)+\frac{\tilde\gamma}2} V_{- r }
(\tau-\gamma+\tilde\gamma)
V_{\frac1\theta- r }(\tau-\gamma-\tilde\gamma),   \\
d_1^\ci(\tau)&=& - e^{\frac\tau2+\gamma(\frac12+\theta- r 
\theta)+\frac{\tilde\gamma}2} V_{1- r  }(\tau-\gamma+\tilde\gamma)
V_{1+\frac1\theta- r  }(\tau-\gamma-\tilde\gamma),
\end{array}
\label{solAB}
\end{equation}
where   the function $V_ r  (\tau)$ is defined by
\begin{equation}\la{intrv}
 \log V_ r  (\tau) ~\defeq~ -\frac12\int\frac{d\omega}\omega\left[
 \frac{e^{-i\omega\tau}\sinh(\pi r  \omega)}
 {\sinh(\pi\omega)\sinh\frac{\pi\omega}\theta} -\frac{ r 
 \theta}{\pi\omega}\right]\, .\quad 
\end{equation}
The properties of the function $V_r(\t)$ are listed in Appendix \ref{sec:Vrt}.

The solution (\ref{solAB}) reproduces correctly the scaling exponents
(\ref{scalingd}) and it is unique, assuming that the correlators
$d_0^\ci$ and $d_1^\ci$ have no poles as functions of $x$.  Near the
branch $AS_\bu$, the functions $d_0 ^\bu$ and $d_1^\bu$ are given by
the same expressions \re{solAB}, but with $ r $ replaced by $ 1/\th -
 r $.

\subsection{Analysis of the solution}

To explore the scaling regimes of the solution \re{solAB} we use the
expansion \re{expV} and return to the original variables,
  \be e^\t=x, \quad e^{\g \th}=(-t)^{1\over\th}, \quad e^{(\g\pm
  \tilde\g)\th} =\( \mp {{t_B\over 2} +\sqrt{{t_B^2 \over
  4}-t}}\)^{1\over\th} .  \ee
Let us define the function $\hat V(x) $ by $V_r(\t) = \hat V_r(e^\t)
$.  The large $x$ expansion of $\hat V_r$ goes, according to
\re{Vassime}, as
\be\la{asVr} \hat V_r(x)= x^{ r \th/2} \( 1 + {\sin\pi  r \over \sin
\pi/\th} \ x^{-1} + {\sin\pi  r  \th \over \sin \pi \th} \ x^{-\th}
+\dots\). \ee 
The expansion at small $x$ follows from the symmetry
$\hat V(x)= \hat V(1/x)$.  Written in terms of the original variables,
the scaling solution near the branch $AS_\ci$ takes the form
   \be
   d_0^\ci (x)&= &
      {1\over w_0} \  {(-t)^{-{r\over 2}}\over \sqrt{x} } \
       \(-{t_B/2  } + \sqrt{{t_B^2/ 4 }-t}
   \)^{{1\over 2\th}  }   \no\\
   &\times&
    \hat V_{ -r}\[  x  \,
   \( {t_B/ 2 } +\sqrt{{t_B^2/ 4 }-t}
   \)^{-{1\over\th}}\]
    \hat V_{{1\over\th} -r}\[ x
   \(-{t_B/ 2 } + \sqrt{{t_B^2/ 4 }-t}
   \)^{-{1\over\th}}\]
   \la{d0sol}
    \\
      d_1^\ci (x)&=& -\sqrt{x}\,
         t^{1-r \over 2}
    \(-{t_B/ 2}+\sqrt{{t_B^2/ 4 }-t}\)^{{1\over 2\th}    } 
    \no\\
 &\times&  \!\!\hat V_{1 -r}\!\!\[  x  \,
   \(  {t_B/ 2 } + \sqrt{{t_B^2/ 4 }-t}
   \)^{-{1\over\th}}\]
    \hat V_{1+ {1\over \th}-r}\!\!\[ x
   \(-{t_B/ 2  } + \sqrt{{t_B^2/ 4 }-t}
   \)^{-{1\over\th}}\].
   \quad
   \la{d1sol}
   \ee

The critical regimes of this solution are associated with the limits
$t\to -0, -\infty$ and $t_B\to 0, \pm\infty$ of the bulk and the
boundary temperature couplings.

\bigskip
 {\it (i)\ Dilute phase, anisotropic special transitions }
\smallskip

This critical regime is achieved when both $t$ and $t_B$ are small.
Using the asymptotics \re{asVr}, we find that in the limit $(t_B\to 0,
t\to - 0)$ the expressions \re{d0sol}-\re{d1sol} reproduce the correct
scaling exponents \re{scalingd} in the dilute phase:
 \be AS_\ci: \qquad d_0^\ci \sim x^{- r  \th} , \qquad d_1^\ci\sim x^{
 1+ \th -\th  r  } \qquad\qquad ( t_B= 0, \ t\to - 0) .  \ee
The regime $AS_\bu$ is obtained by replacing $\ci\to\bu,\  r \to
{1\over\th}- r $.

\bigskip
  {\ \it (ii) Dilute phase,   ordinary transition}
\smallskip

At $t\to-0$, the leading behavior of $ d_0^\bu = 1/d_0^\ci$ and $
d_1^\ci$ for large $t_B$ is (we omitted all numerical coefficients)
 \begin{equation}
\begin{array}{lll}
   d_0^\bu &\sim & {t_B}^{  r }   \(
    1 + {t_B}^{ - \frac1\theta}x   + t_B^{-1} \, x^{ \th } +  \dots \)  
   \\
      d_1^\ci &\sim &  {t_B}^{ 1- r  }
   \(
    x  + {t_B}^{ - \frac1\theta}x^2  + t_B^{-1} \, x^{ \th +1} +
  t_B^{-2} \, x^{2 \th +1} +  \dots \)  
  \end{array}  \quad (t\to- 0,\ t_B\to +\infty) . 
   \label{d0d1p}
  \end{equation}
 In the expansion for $d_0^\bu$, the first singular term, $x^\th$, is
 the singular part of $D_0$ with ordinary/ordinary boundary
 conditions.  In the expansion for $d_1^\ci$, the first singular term,
 $x^{1+\th}$, is the one-point function $w$, while the next term,
 $x^{2\th+1}$, is the singular part of the boundary two-point function
 $D_1^\ci$ with ordinary/ordinary boundary conditions.

\bigskip
  {\ \it (iii) Dilute phase,   extraordinary transition}
\smallskip

 Now we write the asymptotics of \re{d0sol}-\re{d1sol} in the opposite
 limit, $t\to-0$, and $t_B\to-\infty$:
\begin{equation}
\begin{array}{lll}
   d_0^\bu &\sim & {t_B}^{- {1\over\th}+ r }   \(
    x  + {t_B}^{ - \frac1\theta}x^2  + t_B^{-1} \, x^{ \th +1} +
  t_B^{-2} \, x^{2 \th +1} +  \dots \)  
   \\
      d_1^\ci &\sim &  {t_B}^{{1\over\th} +1- r  }
  \(1+  {t_B}^{ - \frac1\theta} \,   x\,
  +   \, t_B^{-1} \, x^{ \th} +\dots \)  
  \end{array}  \qquad (t= 0,\ t_B\to -\infty) .
  \label{d0d1m}
  \end{equation}
This asymptotics reflects the symmetry of  the solution (\ref{solAB})
which maps
 \be\la{sim01}
 r  \to
  1+1/\theta- r  ,   \qquad  d_0^\bu \leftrightarrow  d_1^\ci 
  \ee
which is also a symmetry of the loop equations \re{D0D1}.  In the
limit of large and negative $t_B$, the function $d_0^\bu$ behaves as
the singular part of the correlator $D_1^\ci$ with ordinary/ordinary
boundary conditions, while $d_1^\ci$ behaves as the singular part of
the correlator $D_0$ with ordinary/ordinary boundary conditions.

The asymptotics of the solution at $t_B\to\pm\infty$ confirms the
qualitative picture proposed in \cite{DJHS2} and explained in the
Introduction.  When $t_B$ is large and positive, the loops avoid the
boundary  and we have the ordinary boundary condition.
In the opposite limit, $t_B\to-\infty$, the \DJS  boundary tends to be
coated by loop(s).  Therefore the typical loop configurations for
$D_1^\ci$ in the limit $t_B\to-\infty$ will look like those of
$D_0^\ci$ in the ordinary phase, because the open line connecting the
two boundary-changing points will be adsorbed by the \DJS  boundary.
Conversely, the typical loop configurations for $D_0^\ci$ will look
like those of $D_1^\ci$ in the ordinary phase, because free part of
the loop that wraps the \DJS  boundary will behave as an open line
connecting the two boundary-changing points.

We saw that the solution reproduces the qualitative phase diagram for
the dilute phase, shown in Fig. \ref{fig:FhaseDR}.   Now let us try to
reconstruct the phase diagram in the dense phase.

\bigskip
  {\ \it (iv) Dense phase,   anisotropic special  transitions}
\smallskip

For any finite value of $t_B$, the dense phase is obtained in the
limit $t\to-\infty$.  The asymptotics of \re{d0sol}-\re{d1sol} in this
limit does not depend on $t_B$:
\be d_0^\ci \sim x^{  r  \th-1} \, (-t)^{{1\over 2\th}- r }, \ \quad
d_1^\ci \sim x^{- (1- r  ) \th}\ (-t)^{1-  r  +{1\over 2\th} } \qquad
( t\to -\infty) .  \la{asd0d} \ee
This means that in the dense phase the \DJS  boundary condition is
automatically conformal for any value of $t_B$.  The boundary critical
behavior does not change with the isotropic boundary coupling $t_B$,
but it can depend on the anisotropic coupling $\Delta$.  The solution
\re{d0sol}-\re{d1sol} holds for any positive value of $\Delta$.  For
negative $\Delta$ we have another solution, which is obtained by
replacing $\ci\to \bu$ and $r\to 1/\th - r$.  Thus in the dense phase
there are two possible critical regimes for the \DJS  boundary, one for
positive $\Delta$ and the other for negative $\Delta$, which are
analogous to the two anisotropic special transitions in the dilute
phase.  The domains of the two regimes are separated by the isotropic
line $\Delta=0$.

The above is  true when  $t_B$ is finite.  If $t_B$ tends to
$\pm\infty$, we can obtain critical regimes with the
properties of the ordinary and the extraordinary transitions.

  \bigskip {\ \it (v) Dense phase, ordinary and extraordinary
  transitions } \smallskip

If we expand the solution \re{d0sol}-\re{d1sol} for $- t\gg x^{2\th}$
and $t_B \gg - t x^{-\th}$, the singular parts of the two correlators
will be the same as the correlators with ordinary boundary condition
on both sides.  For example, instead of the term $t_B^{-1} x^\th$ in
the expression for $d_0^\bu$, we will obtain $t^{-1} t_B x^{-\th}$.
This is the singular part of the correlator $D_0$ with ordinary
boundary conditions in the dense phase.  Further, the asymptotics of
the solution in the limit $- t\gg x^{2\th}$ and $t_B \ll t x^{-\th}$
is determined by the symmetry \re{sim01}.  This critical regime has
the properties of the extraordinary transition, in complete analogy
with the dilute case.  We conclude that the ordinary and the
extraordinary transitions exist also in the dense phase, but they are
pushed to $t_B\to\pm\infty$.
       
Finally, let us comment on the possible origin of the square-root
singularity of the solution \re{d0sol}-\re{d1sol} at $t_B^2=4t$.  This
singularity appears in the disordered phase, $t>0$, which is outside
the domain of validity of the solution.  Nevertheless, one can
speculate that this singularity is related to the surface transition,
which separates the phases with ordered and disordered spins near the
\DJS  boundary.  The singularity in our solution has two branches, $t_B=
\pm 2\sqrt{t}$, while in the true solution for $t>0$ the negative
branch should disappear.

\section{Conclusions}
\label{sec:Concl}

In this paper we studied the dilute boundary $O(n)$ model with a class
of anisotropic boundary conditions, using the methods of 2D quantum
gravity.  The loop gas formulation of the anisotropic boundary
conditions, proposed by Dubail, Jacobsen and Saleur (\DJS ), involves two
kinds of loops having fugacities $\nci$ and $\nbu= n-\nci$.  Besides
the bulk temperature, which controls the length of the loops, the
model involves two boundary coupling constants, which define the
interaction of the two kinds of loops with the boundary.  
%The two-dimensional space of boundary couplings is spanned by the
%isotropic and the anisotropic axes.

The regime where the \DJS boundary condition becomes conformal
invariant is named in \cite{DJHS2} anisotropic special transition.
The enhanced symmetry of the model after coupling to gravity system
allowed us to solve the model analytically away from the anisotropic
special transition.  We used the solution to explore the deformations
away from criticality which are generated by the bulk and boundary
thermal operators.

Our main results can be summarized as follows.

1) We found the phase diagram for the boundary transitions in the
dilute phase of the $O(n)$ model with anisotropic boundary
interaction.  The phase diagram is qualitatively the same as the one
obtained in \cite{DJHS2} and sketched in the Introduction.  The
critical line consists of two branches placed above and below the
isotropic line.  Near the special point the critical curve is given by
the same equation on both sides of the isotropic line, which means
that the two branches of the critical line meet at the special point
without forming a cusp.  We also demonstrated that the analytic shape
of the critical curve does not contradict the scaling of the two
boundary coupling constants.  This contradicts the picture drawn in
\cite{DiEi} on the basis of scaling arguments, which seems to be
supported by the numerical analysis of \cite{DJHS2}.  Of course we do
not exclude the possibility that the origin of the discrepancy is in
the fluctuations of the metric.

2) From the singular behavior of the boundary two-point functions we
obtained the spectrum of conformal dimensions of the $L$-leg boundary
operators between ordinary and anisotropic special boundary conditions,
which is in agreement with \cite{DJHS2}.  In order to establish the
critical exponents we used substantially the assumption that $\nci$
and $\nbu$ are both non-negative.
   
3) We showed that the two-point functions of these operators coincide
with the two-point functions in boundary Liouville theory \cite{FZZb}.
The functional equation for the boundary two-point function obtained
from the Ward identities in the matrix model is identical to the
functional equation derived by using the OPE in boundary Liouville
theory.

4) The result which we find the most interesting is the expression for
the two-point functions away from the critical lines.  For any finite
value of the anisotropic coupling $\Delta $, the deviation from the
critical line is measured by the renormalized bulk and the boundary
thermal couplings, $t$ and $t_B$.  Our result, given by eqs.
\re{d0sol}-\re{d1sol}, gives the boundary two-point function in a
theory which is similar to boundary Liouville gravity, except that the
bulk and the boundary Liouville interactions are replaced by the
Liouville dressed bulk and boundary thermal operators, $t\, \CO_{1,3}$
and $t_B\, \CO_{1,3}^B$.  The boundary flow, generated by the boundary
operator $\CO_{1,3}^B$, relates the anisotropic special transition
with the ordinary and the extraordinary ones.  The bulk thermal flow,
generated by the operator $\CO_{1,3}$, relates the dilute and the
dense phases of the $O(n)$ model coupled to gravity.  At the critical
value of the boundary coupling, the bulk flow induces a boundary flow
between one \DJS boundary condition in the dilute phase and another
\DJS boundary condition in the dense phase.  For the rational values
of the central charge, the boundary conditions associated with the
endpoints of the bulk flow match with those predicted by the recent
study using perturbative RG techniques \cite{Fredenhagen:2009sw}.

Here we considered only the boundary two-point functions with
ordinary/DJS boundary conditions.  It is not difficult to write the
loop equations for the boundary $(n+1)$-point functions with one
ordinary and $n$ \DJS boundaries.  The loop equations for $n>1$ will
depend not only on the parameters characterizing each segment of the
boundary, but also on a hierarchy of overlap parameters that define
the fugacities of loops that touch several boundary segments.  The
loop equations for the case $n=2$ were studied for the dense phase in
\cite{Bourgine:2009hv}.  In this case there is one extra parameter,
associated with the loops that touch both \DJS boundaries, which
determines the spectrum of the boundary operators compatible with the
two \DJS boundary conditions.  In the conformal limit, the loop
equations for the $(1+n)$-point functions should turn to boundary
ground ring identities, which, compared to those derived in
\cite{Kostov:2004cq} for gaussian matter field, will contain a number 
of extra contact terms with coefficients determined by the overlap parameters.
It would be interesting to generalize the calculation of  \cite{Bourgine:2009hv} 
to the dilute case and compare with the existing results 
\cite{Basu:2005sda, Furlan:2008za} for the 3-point functions in 
 Liouville gravity with  non-trivial matter field.

The method developed in this paper can be generalized in several
directions.  Our results were obtained for the $O(n)$ model, but they
can be easily extended to other loop models, as the dilute ADE hight
models.  It is also clear that the method works for more general cases
of anisotropic boundary conditions, with the $O(n)$ invariance broken
to $O(n_1)\times\dots \times O(n_k)$.

Finally, let us mention that there is an open problem in our approach.
The loop equations does not allow to evaluate the one-point function
with \DJS boundary conditions, $H(y)$, except in some particular
cases.  There are two possible scaling limits for this function, which
correspond to the two Liouville dressings of the identity operator
with \DJS boundary condition, and it can happen that both dressings
are realized depending on the boundary parameters.  This ambiguity
does not affect the results reported in this paper.

\section*{Acknowledgments}

 We thank J. Dubail, J. Jacobsen and H. Saleur for useful discussions.
 This work has been supported in part by Grant-in-Aid for Creative
 Scientific Research (project \#19GS0219) from MEXT, Japan, European
 Network ENRAGE (contract MRTN-CT-2004-005616) and the ANR program
 GIMP (contract ANR-05-BLAN-0029-01).

\appendix

%%%%%%%%%%%%%%%%%%%%%%%%%%%%%%%%%%%%%%%%%%
\section{ Derivation of the loop equations}
%%%%%%%%%%%%%%%%%%%%%%%%%%%%%%%%%%%%%%%%%%
\label{sec:Derivation}

Here we give the derivation of the loop equations which are extensively
discussed in this work.

We first summarize our technique to derive loop equations.  The
translation invariance of the matrix measure implies, for any matrix
${\bf F}$ made out of ${\bf X}$ and ${\bf Y}_a$, the identities
\begin{eqnarray}
\label{WE1}\frac1\beta \langle\partial_{\bf X}{\bf F}\rangle + \langle
\text{tr}\big[(-V'({\bf X})+\vec{\bf Y}^2){\bf F}\big] \rangle &=& 0,
\\
\label{WE2} \frac1\beta \langle \partial_{{\bf Y}_a}{\bf F}\rangle
+\langle \text{tr}\big[{\bf Y}_a({\bf XF+FX})\big] \rangle &=&0 \,,
\end{eqnarray}
where the derivatives with respect to matrices are defined by
$\partial_{\bf X}{\bf F}\equiv \partial F_{ij}/\partial X_{ij}$ summed
over the indices $i$ and $j$, and are generally given by sums of
double traces.  Written for the observable ${\bf G}= - ({\bf XF+FX})$,
the second equation states\footnote{ Here it is assumed that the
eigenvalues of ${\bf X}$ are all in the left half plane.  This is
indeed true in the large $N$ limit of our matrix integral.  }
\begin{equation}\label{nonlocalWE}
 \left\langle\text{tr}\big({\bf Y}_a{\bf G}\big)\right\rangle =
 \frac1\beta \oint_{i\mathbb{R}}\frac{dx}{2\pi
 i}\left\langle\partial_{{\bf Y}_a} \big[{\bf W}(x){\bf
 GW}(-x)\big]\right\rangle .
\end{equation}
These identities are used along with the large $N$ factorization
\begin{equation}
 \left\langle\text{tr}{\bf A}\cdot\text{tr}{\bf B}\right\rangle \simeq
 \left\langle\text{tr}{\bf A}\right\rangle \left\langle\text{tr}{\bf
 B}\right\rangle
\end{equation}
to derive various relations among disk correlators.

%%%%%%%%%%%%%%%%%%%%%%%%%%%%%%%%%%% 
\subsection{Loop equation for the resolvent}
%%%%%%%%%%%%%%%%%%%%%%%%%%%%%%%%%%%% 

An equation for the resolvent (\ref{resW}) is obtained if we take
${\bf F}={\bf W}(x)$. The identity (\ref{WE1}) then gives
\begin{equation}\label{WE1bis}
 W^2(x) = \frac1\beta
 \left\langle \text{tr}\big[V'({\bf X}){\bf W}(x)\big]\right\rangle
 -\sum_{a=1}^n\frac1\beta
 \left\langle\text{tr}\big[{\bf Y}_a^2{\bf W}(x)\big]\right\rangle.
\end{equation}
Then the identity (\ref{nonlocalWE}) applied to the last term gives
\begin{eqnarray}
 \left\langle\text{tr}\big[{\bf Y}_a^2{\bf W}(x)\big]\right\rangle &=&
 \frac1\beta\oint_{i\mathbb{R}} \frac{dx'}{2\pi i}
 \left\langle\partial_{{\bf Y}_a} \big[{\bf W}(x'){\bf W}(x){\bf
 Y}_a{\bf W}(-x')\big]\right\rangle \nonumber \\
&=& \frac1\beta \oint_{i\mathbb{R}}\frac{dx'}{2\pi i}
\left\langle\text{tr}\big[{\bf W}(x'){\bf W}(x)\big]
\text{tr}\big[{\bf W}(-x')\big]\right\rangle \nonumber \\
 &=& -\beta\int_{i\mathbb{R}}\frac{dx'}{2\pi i}\frac{W(x)-W(x')}{x-x'}W(-x').
\label{eqnY}
\end{eqnarray}
 Using the $\star$-product introduced in (\ref{defstar}),
 the last line can be written as $\beta[W\star W](x)$.   The equation
(\ref{WE1bis}) can then be written as
\begin{equation}\label{eqnW}
 W(x)^2-V'(x)W(x)+ n[W\star W](x)= \frac1\beta\left\langle\text{tr}
\Big(\frac{V'({\bf X})-V'(x)}{x-{\bf X}}\Big)\right\rangle.
\end{equation}
For a cubic potential the expectation value on the r.h.s. is a
polynomial of degree one.  Using an important property of the
$\star$-product
\begin{equation}\label{starsym}
 [f\star g ](x)+[g\star f](-x) = f(x)g(-x)\,,
\end{equation}
which can be proved by deforming the contour of integration, one
obtains a loop equation \cite{Kostov:1991cg}, which is a quadratic
functional equation for $W(x)$.  The term linear in $W$ can be
eliminated by a shift
\begin{equation}
 w(x)~\defeq~ W(x)-\frac{2V'(x)-nV'(-x)}{4-n^2}.
\end{equation}
The loop equation for $w(x)$ is given by (\ref{funceqW}).

 %%%%%%%%%%%%%%%%%%%%%%%%%%%%%%%%%%%%%%  
\subsection{Loop equations for the boundary two-point functions}
%%%%%%%%%%%%%%%%%%%%%%%%%%%%%%%%%%%%% 

The boundary two-point functions of $L$-leg operators satisfy the
recurrence equations
$$
 D_{L+1}^\ci = W\star D_L^\ci\,,\quad
 D_{L+1}^\bu = W\star D_L^\bu\,\quad(L\ge1)
$$
which can be derived by applying (\ref{nonlocalWE}) with ${\bf F}={\bf
W}{\bf S}_{L+1}{\bf H}{\bf S}_L$.  By applying (\ref{nonlocalWE}) to
${\bf F}={\bf W}{\bf Y}_a{\bf H}$ and ${\bf W}{\bf H}{\bf Y}_a$, we
find
\begin{eqnarray}
 D_1^\ci &=& (\mci  D_1^\ci+ W)\star D_0, \nonumber \\
 \frac{1}{\beta\nci}\left\langle\text{tr}\big[{\bf
 WHY}_\ci^2\big]\right\rangle &=& D_0\star (\mci  D_1^\ci+ W),
\label{D0D1-Y}
\end{eqnarray}
and a similar pair of equations for $D_0$ and $D_1^\bu$.  The first
equation of (\ref{D0D1-Y}) can be rewritten into an equation for the
discontinuity along the branch cut,
\begin{equation}
\text{Disc}D_1^\ci(x)~=~D_0(-x)\text{Disc}(\mci  D_1^\ci(x)+  W(x)),
\end{equation}
which implies that the recurrence equation can be extended to $L=0$ by
defining
$$
 D_0^\ci\defeq \frac{  D_0}{\mci  D_0-1},\qquad
 D_0^\bu\defeq \frac{ D_0}{\mbu D_0-1}.
$$
Also, by applying (\ref{WE1}) to ${\bf F}={\bf WH}$ one finds
\begin{equation}
 (W+H)D_0 ~=~ V'(x)D_0-P(x) -\frac1\beta\left\langle\text{tr}
 \big[{\bf WH}
 \sum_{\a=1,2}
  {\bf Y}_\aa^2
  \big]\right\rangle\,,
\label{D0D1-X}
\end{equation}
where $P(x)$ is defined in (\ref{defP}).

By using (\ref{starsym}) to combine the two equations in
(\ref{D0D1-Y}) and noticing that
$$
 \text{tr}\big[{\bf WH}
 \sum_{\a=1,2}
 \maa {\bf Y}_\aa^2 ] ~=~
 \text{tr}\big[{\bf H}+(y-x){\bf WH}-{\bf W}],
$$
one finds a quadratic relation
\begin{eqnarray}
 \lefteqn{\la{KFEQ1b} (y-x)D_0(x)+H-W(x)+
 \sum_{\a=1,2}
  \maa  \naa D_1^\aa(-x) } \nonumber \\ &=& 
  D_0(x)
  \sum_{\a=1,2}
  \(\maa ^2\naa
 D_1^\aa(-x) +\maa  \naa W(-x) \)
 .
\end{eqnarray}
By inserting the second of the equation (\ref{D0D1-Y}) into
(\ref{D0D1-X}) one finds another quadratic equation,
\begin{eqnarray}\la{KFEQ2b}
 \lefteqn{(W(x)+H)D_0(x)+P(x)-V'(x)D_0(x)-
  \sum_{\a=1,2}
 \naa D_1^\aa(-x) }
 \nonumber\\ &=& -D_0(x)\Big\{
 \sum_{\a=1,2}\maa  \naa
 D_1^\aa(-x) +nW(-x)\Big\}.
\end{eqnarray}
These two equations 
are equivalent to \re{KFEQ1}  and \re{KFEQ2}.

%%%%%%%%%%%%%%%%%%%%%%%%%%%%%%
\section{2D Liouville gravity}
%%%%%%%%%%%%%%%%%%%%%%%%%%%%%%
\label{sec:Liouv}

In 2D Liouville gravity formalism, the $c\le 1$ matter CFTs are
coupled to the Liouville theory with central charge $26-c$ and the
reparametrization ghosts.  For Liouville theory, we denote the standard
coupling by $b$ and the background charge by $Q=b+1/b$.  The central
charge is given by $26-c=1+6Q^2$.  In our convention, $b$ is always
smaller than one. When the matter CFT is $(p,q)$ minimal model, we have
\begin{equation}
  b=\sqrt{\frac pq},\qquad c=1-\frac{6(p-q)^2}{pq}.
\end{equation}
In studying the $O(n)$ model we used the parametrization
$n=2\cos\pi\theta$.  If $\theta=1/p$ with $p\in\mathbb{Z}$, the model
describes the flow between $(p,p+1)$ and $(p-1,p)$ minimal models
corresponding respectively to the dilute and dense phase critical
points.

The primary operators in $(p,q)$ minimal model fit in the Kac table
which has $(p-1)$ rows and $(q-1)$ columns.  The operator $\Phi_{r,s}$
has the conformal dimension
\begin{equation}
h _{r,s}=\frac{(rq-sp)^2-(q-p)^2}{4pq}
 =\frac{(r/b-sb)^2-(1/b-b)^2}4,
\end{equation}
and are subject to the identification $\Phi_{r,s}=\Phi_{p-r,q-s}$.  In
2D Liouville gravity coupled to the $(p,q)$ minimal CFT, the operator
$\Phi_{r,s}$ is dressed by the Liouville exponential
$e^{2\alpha_{r,s}\phi}$ or $e^{\alpha_{r,s}\phi}$ depending on whether
it is a bulk or boundary operator, so that the total conformal weight
becomes one.  This requires
\begin{equation}
 \alpha_{r,s}(Q-\alpha_{r,s})+h _{r,s}=1,\qquad
 2\alpha_{r,s}= Q\pm\big(r/b-sb\big).
\label{alpha}
\end{equation}

\subsection{Conformal weight and scaling exponents of couplings}

In making the comparison between the matrix model and Liouville
gravity, we start from the fact that the resolvent $w(x)$ and its
argument $x$ correspond to the two boundary cosmological constants
$\mu _B$ and $\tilde\mu _B$.  They couple respectively to the
boundary cosmological operators $e^{b\phi}$ and $e^{\phi/b}$, and are
therefore proportional to $\mu ^{\frac12}$ and
$\mu ^{\frac1{2b^2}}$, where $\mu $ is the Liouville bulk
cosmological constant.  Since $w(x)\sim x^{1\pm\theta}$ in the dilute
and dense phases, we find
\begin{equation}
\begin{array}{llll}
 \text{(dilute):} &
 x=\mu _B,&
 w(x)\sim x^{1+\theta}=\tilde\mu _B,&
 b=(1+\theta)^{-1/2}, \\
 \text{(dense):} & x=\tilde\mu _B,& w(x)\sim
 x^{1-\theta}=\mu _B,& b=(1-\theta)^{+1/2}.
\end{array}
\end{equation}

Suppose a boundary condition has a deformation parametrized by a
coupling which scales like $x^\rho$.  Then the corresponding boundary
operator should be dressed by the Liouville operator with momentum
$\alpha=\rho b$ in the dilute phase and $\alpha=\rho/b$ in the dense
phase.  Then using (\ref{alpha}) one can determine the conformal
dimension of the operator responsible for the boundary deformation.
Let us apply this idea to the deformations parametrized by $\mu_B$,
$t_B$ and $\Delta$ in Sect. \ref{sec:Scalimit}.

\paragraph{Deformation by $\mu_B$.} $\mu_B$ scales like $x^{1+\theta}$
in the dilute phase and $x^{1-\theta}$ in the dense phase.  So the
corresponding operators are dressed by
$e^{(1+\theta)b\phi}=e^{\phi/b}$ in the dilute phase and
$e^{(1-\theta)\phi/b}=e^{b\phi}$ in the dense phase.  The matter
conformal dimension is zero, i.e., the operator responsible for the
deformation is the identity $\Phi_{1,1}$.

\paragraph{Deformation by $t_B$.} $t_B$ scale like $x^\theta$ in the
dilute phase and $x^{-\theta}$ in the dense phase, so the
corresponding operator gets dressings with Liouville momentum
$b\theta=\frac1b-b$ in the dilute phase and $-\theta/b=b-\frac1b$ in
the dense phase.  We identify the operator with $\Phi_{1,3}$
(relevant) in the dilute phase and $\Phi_{3,1}$ (irrelevant) in the
dense phase.

\paragraph{Deformation by the anisotropy coupling $\Delta$.} The coupling
scales like $x^{1-\theta}$ in the dilute phase.  The corresponding
operator is dressed by the Liouville momentum $b(1-\theta)=2b-\frac1b$
and therefore identified with $\Phi_{3,3}$.

\subsection{Conformal weight and scaling exponents of correlators}

After turning on the gravity, correlators no longer depend on the
positions of the operators inserted because one has to integrate over
the positions of those operators.  The dimensions of the operators
therefore should then be read off from the dependence of correlators
on the cosmological constant $\mu$.  If we restrict to disk
worldsheets, the amplitudes with $n$ boundary operators $\CO ^B_i$
dressed by Liouville operators $e^{\beta_i\phi}$ and $\l $  bulk
operators $\CO _i$ dressed by $e^{2\alpha_j\phi}$ scale with $\mu$ as
\begin{equation}
 \big\langle \CO ^B_1\cdots\CO ^B_n\CO _1\cdots\CO _m
 \big\rangle \propto \mu ^{\frac{1}{2b}(Q-2\sum\alpha_j-\sum\beta_i)}.
\label{mu-col}
\end{equation}

Suppose a disk two-point function of a boundary operator $\CO ^B$
scales like $x^\rho$.  Then $\CO ^B$ should be dressed with Liouville
momentum $\beta=\frac12(Q-b\rho)$ in the dilute phase and
$\beta=\frac12(Q-\rho/b)$ in the dense phase.  We thus identify
$\CO ^B$ with the $(r,s)$ operator in the Kac table if
\begin{equation}
\pm\rho=r(1+\theta)-s ~~~(\text{dilute phase}),\qquad
\pm\rho=r-s(1-\theta) ~~~(\text{dense phase}).
\end{equation}

\subsection{Boundary two-point function}

In Liouville theory with FZZT boundaries, the disk two-point structure
constant is
\begin{equation}
 d(\beta|t,s) =
 \frac{(\mu \pi\gamma(b^2)b^{2-2b^2})^{(Q-2\beta)/2b} {\bf
 G}(Q-2\beta){\bf G}^{-1}(2\beta-Q)} {{\bf S}(\beta+it+is){\bf
 S}(\beta+it-is) {\bf S}(\beta-it+is){\bf S}(\beta-it-is)}.
\end{equation}
Here the special functions ${\bf G}$ and ${\bf S}$ satisfy
\begin{equation}
 {\bf S}(x+b)=2\sin(\pi b x){\bf S}(x),~~~~~ {\bf
 G}(x+b)=\frac{1}{\sqrt{2\pi}}b^{\frac12-b x}\Gamma(b x){\bf G}(x),
\end{equation}
and similar equations with $b$ replaced by $1/b$.  The boundary
parameter $s$ is related to the boundary cosmological constant by
\begin{equation}
 \mu _B=\Big(\frac{\mu }{\sin\pi b^2}\Big)^\frac12\cosh(2\pi b
 s),~~~~ \tilde\mu _B=\Big(\frac{\tilde\mu }{\sin\pi
 b^{-2}}\Big)^{\frac12} \cosh(2\pi s/b).
\end{equation}
where $\mu $ is the bulk cosmological constant and $\tilde\mu $
is its dual, \be (\mu \pi\gamma(b^2)b^{2-2b^2})^{1/b}
=(\tilde\mu \pi\gamma(b^{-2})b^{-2+2b^{-2}})^{b}.  \ee Our loop
equation \re{D0D1} can be compared with
$$
 d(\beta|t,s)d(Q-\beta-\tfrac b2|t\pm\tfrac{ib}2,s) ~=~ F(\beta)
 \Big\{\cosh 2\pi b(t\pm i\beta)-\cosh 2\pi bs\Big\},
$$
\begin{equation}
 F(\beta) ~\defeq~ \sqrt{\frac\mu {\sin\pi b^2}}
 \frac{\Gamma(2b\beta-b^2-1)\Gamma(1-2b\beta)}{\Gamma(-b^2)}.
\end{equation}
or the one with $b$ replaced by $1/b$.  Similarly, the recurrence
equation (\ref{recestar}) among disk correlators of $L$-leg operators
can be compared with
$$
 \frac{d(\beta|t+\frac{ib}2,s)-d(\beta|t-\frac{ib}2,s)} {d(\beta+\frac
 b2|t,s)} ~=~ G(\beta)\Big\{ \cos 2\pi ib(t+\tfrac{ib}2) -\cos 2\pi
 ib(t-\tfrac{ib}2) \Big\},
$$
\begin{equation}
 G(\beta)~\defeq~ -\sqrt{\frac\mu {\sin\pi b^2}}
 \frac{\Gamma(1+b^2)\Gamma(1-2b\beta)}{\Gamma(2+b^2-2b\beta)}.
\end{equation}
%

   %%%%%%%%%%%%%%%%%%%%%%%%%%%%%%
\section{Properties of the function $V_r(\t)$}
%%%%%%%%%%%%%%%%%%%%%%%%%%%%%%
\label{sec:Vrt}

 The function  $V_r(\t)$
   defined by 
   \begin{equation}\la{intrvA}
 \log V_ r  (\tau) ~\defeq~ -\frac12\int_{-\infty}^\infty
 \frac{d\omega}\omega\left[
 \frac{e^{-i\omega\tau}\sinh(\pi r  \omega)}
 {\sinh(\pi\omega)\sinh\frac{\pi\omega}\theta} -\frac{ r 
 \theta}{\pi\omega}\right]\, 
\end{equation}
  satisfies the shift relations
\begin{eqnarray}\la{expV}
 V_{ r  +1}(\tau) &=& 2\cosh\Big(\frac{\theta(\tau\pm i\pi r 
 )}2\Big)V_ r  (\tau\mp i\pi), \nonumber \\
 V_{ r  +\frac1\theta}(\tau) &=& 2\cosh\Big(\frac{\tau\pm i\pi r 
 }2\Big)V_ r  (\tau\mp \tfrac{i\pi}\theta),
\end{eqnarray}
which follow from the integration formula
\be
 \log(2\cosh t)~=~\int_{-\infty}^\infty\frac{d\omega}{2\omega}
 \left[-\frac{e^{-2i\omega t}}{\sinh(\pi\omega)}+\frac1{\pi\omega}\right].
\ee

By deforming the contour of integration and applying the Cauchy
theorem we can write the integral \re{intrvA} as the following formal
series which makes sense for ${{\rm Re}}[\tau]>0$:
\be {\ln V_ r  (\tau)= { \th  r  \over 2}\tau - \sum _{n=1}^{\infty}{
(-)^{ n} \over n} e^{-n\tau} {\sin(n\pi  r  ) \over \sin (\pi n/\th)}
- \sum _{n=1} ^{\infty}{(-)^n \over n} e^{- n \th \tau} { \sin(n \pi
\th r  ) \over \sin (n\pi\th)}.} \ee
The expansion for ${{\rm Re}}[ \t]<0$ follows from the symmetry 
$V_r(\t)=V_r(-\t)$.
The expansion of the function $V_r$ at infinity is
\be \la{Vassime} V_r  (\tau) & =& \ e^{\theta r  |\tau|/2}\ \( 1 +
{\sin(\pi r  ) \over \sin (\pi /\th)} e^{-|\tau|} +{ \sin( \pi \th
r  ) \over \sin ( \pi\th)} \ e^{- \th| \tau|} \ +\dots\) ,\qquad
\tau\to\pm\infty .  \ee

\end{document}